\documentclass[10pt,journal]{IEEEtran}
\IEEEoverridecommandlockouts
\usepackage{cite}
\usepackage{amsmath,amssymb,amsfonts}
\usepackage{algorithm}
\usepackage{algorithmic}
\usepackage{graphicx}
\usepackage{textcomp}
\usepackage{xcolor}
\usepackage{subfigure}
\usepackage[section]{placeins}
\newtheorem{remark}{\text{Remark}}
\newtheorem{lemma}{\text{Lemma}}

\begin{document}

\title{Power-Measurement-Based Channel Autocorrelation Estimation for IRS-Assisted Wideband Communications }
\author{He Sun, \emph{Member, IEEE},
        Lipeng Zhu, \emph{Member, IEEE},
        Weidong Mei, \emph{Member, IEEE},
        and Rui Zhang, \emph{Fellow, IEEE}
\thanks{This work was supported in part by the Advanced Research and Technology Innovation Centre (ARTIC), National University of Singapore under Grant R-261-518-005-720; in part by Guangdong Provincial Key Laboratory of Big Data Computing; in part by the National Natural Science Foundation of China under Grant 62331022; in part by Guangdong Major Project of Basic and Applied Basic Research under Grant 2023B0303000001, and in part by the Natural Science Foundation of Sichuan Province under Grant 2025ZNSFSC0514. Part of this work has been accepted and will be presented at the IEEE Wireless Communications and Networking Conference (WCNC) in 2025, Milan, Italy\cite{sun2024power}. The associate editor coordinating the review of this article and approving it for publication was Prof. M. Sheng (\emph{Corresponding authors}: Rui Zhang; Lipeng Zhu.)} \\
\thanks{H. Sun and L. Zhu are with the Department of Electrical and Computer Engineering, National University of Singapore, Singapore 117583 (e-mail: sunele@nus.edu.sg, zhulp@nus.edu.sg). W. Mei is with National Key Laboratory of Wireless Communications, University of Electronic Science and Technology of China, China 611731 (e-mail: wmei@uestc.edu.cn). R. Zhang is with School of Science and Engineering, Shenzhen Research Institute of Big Data, The Chinese University of Hong Kong, Shenzhen, Guangdong 518172, China (e-mail: rzhang@cuhk.edu.cn). He is also with the Department of Electrical and Computer Engineering, National University of Singapore, Singapore 117583 (e-mail: elezhang@nus.edu.sg).}
}

\markboth{}%
{Shell \MakeLowercase{\textit{et al.}}: Bare Demo of IEEEtran.cls for IEEE Journals}

\maketitle

\begingroup
\allowdisplaybreaks

\begin{abstract}
Channel state information (CSI) is essential to the performance optimization of intelligent reflecting surface (IRS)-aided wireless communication systems. However, the passive and frequency-flat reflection of IRS, as well as the high-dimensional IRS-reflected channels, have posed practical challenges for efficient IRS channel estimation, especially in wideband communication systems with significant multi-path channel delay spread. To tackle the above challenge, we propose a novel neural network (NN)-empowered IRS channel estimation and passive reflection design framework for the wideband orthogonal frequency division multiplexing (OFDM) communication system based only on the user's reference signal received power (RSRP) measurements with time-varying random IRS training reflections. As RSRP is readily accessible in existing communication systems, our proposed channel estimation method does not require additional pilot transmission in IRS-aided wideband communication systems. In particular, we show that the average received signal power over all OFDM subcarriers at the user terminal can be represented as the prediction of a single-layer NN composed of multiple subnetworks with the same structure, such that the autocorrelation matrix of the wideband IRS channel can be recovered as their weights via supervised learning. To exploit the potential sparsity of the channel autocorrelation matrix, a progressive training method is proposed by gradually increasing the number of subnetworks until a desired accuracy is achieved, thus reducing the training complexity. Based on the estimates of IRS channel autocorrelation matrix, the IRS passive reflection is then optimized to maximize the average channel power gain over all subcarriers. Numerical results indicate the effectiveness of the proposed IRS channel autocorrelation matrix estimation and passive reflection design under wideband channels, which can achieve significant performance improvement compared to the existing IRS reflection designs based on user power measurements.
\end{abstract}


\section{Introduction}

The six-generation (6G) mobile communication systems are expected to accommodate substantial emerging wireless applications, by offering unprecedented capabilities such as high spectral efficiency, extremely-low latency, and extensive connectivity\cite{6G,10045774}. Although the massive multiple-input multiple-output (mMIMO) technologies are envisioned to provide enhanced beamforming and spatial multiplexing gains, the increasing demand of radio-frequency (RF) chains in large-scale antenna arrays of active transceivers results in prohibitively high hardware cost and power consumption, which challenge their practical implementation\cite{mMIMO,6GXLMIMO}.
With the recent advances in metamaterials, intelligent reflecting surface (IRS) has evolved as an inventive technology for establishing a cost-effective and adjustable wireless communication environment\cite{wu2023intelligent,mei}. An IRS usually refers to a planar array composed of numerous quasi-passive and low-cost reflecting elements that are able to reflect the incoming signals with adjustable phase shifts/amplitudes. Through collaboratively adjusting the phase shifts/amplitudes of reflecting elements, the electromagnetic signal propagation can be reconfigured to superimpose signals constructively/destructively for enhancing/suppressing the desired/undesired signal at the receiver. Furthermore, different from traditional base stations (BSs)/relays, the passive reflection of IRS eliminates the need for power amplifiers or RF chains, thus significantly reducing the power consumption and hardware expense. Motivated by the above appealing features, the enhancement of IRS performance has been thoroughly researched under different systems, such as MIMO \cite{IRSMIMO1,IRSMIMO2}, secure wireless communication\cite{IRSSecurity1,IRSSecurity2}, simultaneous wireless information and power transfer (SWIPT)\cite{IRSEnergy,IRSEnergy1}, vehicular communication\cite{10316541,MobilityIRS}, target localization\cite{LocIRS1,Peilan}, integrated sensing and communication (ISAC)\cite{IRSISAC2,IRSISAC3,IRSISAC1}, etc.

IRS is expected to make significant contributions to achieving seamless coverage and ubiquitous connectivity in future wireless networks\cite{OFDMIRSCE1}. To effectively harness the potential benefits of IRS for channel reconfiguration, IRS passive reflection/beamforming should be properly designed\cite{Zhengsurvey,mei}. Most of the existing IRS passive reflection designs require channel state information (CSI) or dedicated pilot training. For example, the IRS channels are first estimated by exploiting the received complex-valued pilots, and the IRS reflection patterns are then optimized utilizing the estimated IRS channels for data transmission\cite{OFDMIRSCE1,Zhengsurvey,YCSIRS,CEIRS1,Yangyf,Zheng20201,Zhengbx2020,OFDMIRSCE2,10042005}. Particularly, pilot-based IRS channel estimation methods in wideband orthogonal frequency division multiplexing (OFDM) systems have been extensively studied in the literature. In \cite{OFDMIRSCE1}, a twin-IRS architecture was proposed to estimate channel parameters for wideband communications based on received complex-valued pilots. The authors of \cite{OFDMIRSCE2} proposed a joint pilot and IRS training reflection pattern optimization algorithm to enhance the efficiency of IRS channel estimation in wideband OFDM systems. In \cite{OFDMIRSCE3}, an efficient IRS channel estimation algorithm was proposed to enhance the channel estimation accuracy for IRS-assisted MIMO-OFDM systems by exploiting channel sparsity in both angle and delay domains in the millimeter-wave frequency band. Alternatively, deep learning can be employed for high-dimensional channel estimation in mMIMO or IRS-assisted communication systems \cite{9127834,add001,add002,9775110,YWJSAC,add003}. In these approaches, deep neural networks (DNNs) are trained using large datasets to learn complex inherent channel correlations and leverage them for data-driven channel estimation. To improve the DNN-based channel estimation accuracy, pilot signals and channel estimation methods can be jointly optimized\cite{add001,add002,add003}. In \cite{YWJSAC}, a pilot-based IRS passive reflection design was proposed by exploiting deep learning to predict the optimal IRS passive reflection coefficients based on the received complex-valued pilot signals without explicit IRS channel estimation. However, under the existing wireless system (e.g., cellular or WiFi) protocols, pilot signals are designed only for estimating the BS-user direct channels, without additional resources dedicated to estimating the IRS-associated channels. Thus, traditional pilot-based IRS channel estimation and reflection design requires significant modifications of the current communication system protocols, which thus face a great difficulty in practical implementation. Furthermore, even additional pilots are allowed for IRS channel estimation, the associated channel estimation overhead may be substantial due to the high-dimensional IRS channels.

To facilitate the seamless integration of IRSs into the existing wireless communication systems, recent works \cite{CSM,sunGCC,sunNN1,YGGC} have explored practically implementable IRS passive reflection designs by exploiting the users' received signal (pilot/data) power measurements instead of relying on the complex-valued pilot signals. Since the received power measurements are readily available in modern communication systems, the power-measurement-based IRS reflection design eliminates the need for additional pilot signals, making it seamlessly compatible with existing cellular/WiFi protocols. In particular, the authors in \cite{CSM} proposed a conditional sample mean (CSM) method to design the IRS reflection with discrete phase shifts directly utilizing the reference signal received power (RSRP) samples measured under a set of random reflections of the IRS. It was revealed that given a sufficiently large number of samples, the CSM method can achieve the same receiver signal-to-noise ratio (SNR) scaling order as that under perfect CSI. However, the CSM method requires a large number of user RSRP measurements in general, due to the lack of channel recovery from the power measurements.

To overcome this limitation of CSM, the authors in \cite{sunGCC,sunNN1,YGGC} introduced a new channel recovery approach for the power-measurement-based IRS reflection design. Specifically, given the RSRP measurements under a set of random IRS reflections, they proposed to recover the IRS-cascaded channel by training a single-layer neural network (NN)\cite{sunGCC,sunNN1} or solving an equivalent rank-minimization problem\cite{YGGC}. However, these works only considered frequency-flat narrowband communication systems, while their extensions to the more general frequency-selective wideband channels have not been pursued yet. As compared to narrowband systems, wideband IRS channel recovery is more challenging due to the increased channel coefficients resulting from multi-path channel delay spread and their more intricate effects on the user's RSRP measurements.

\begin{figure}[t]
  \centering
  {\includegraphics[width=0.516\textwidth]{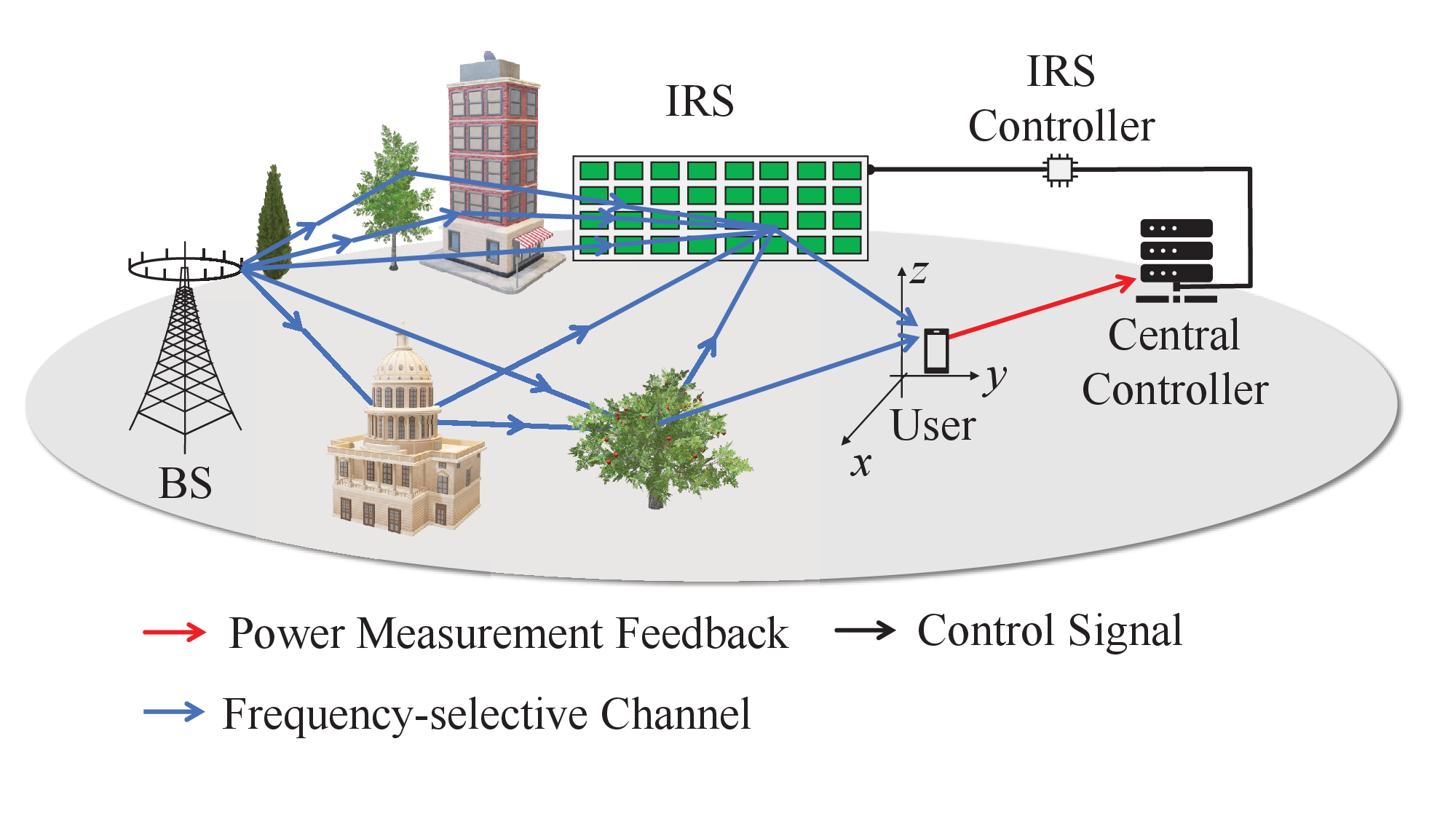}}
  \caption{An IRS-aided wideband OFDM system. }
\label{fig0001}
\end{figure}

To tackle the above challenges, we propose a new user power-/RSRP-measurement-based IRS channel estimation and reflection design framework for the wideband OFDM communication system, as shown in Fig. \ref{fig0001}. The main contributions of this paper are summarized as follows:

\begin{itemize}
  \item First, we derive the average received signal power (ARSP) over all OFDM subcarriers at the user in terms of the time-domain autocorrelation matrix for the IRS-cascaded wideband channel. To estimate this wideband channel autocorrelation matrix, we propose a single-layer NN-enabled method utilizing the user's RSRP measurements over only a subset of subcarriers. Particularly, we reveal that under any given IRS passive reflection, the RSRP can be expressed as the output of a NN, which is the sum of the outputs of multiple single-layer subnetworks, with their number no less than the rank of the autocorrelation matrix of IRS channel. Different from existing pilot-based DNN-enabled  IRS channel estimation methods, the proposed IRS channel estimation method recovers the channel autocorrelation matrix from the weights of the single-layer NN (or its subnetworks) trained via supervised learning.
  \item However, the rank of the channel autocorrelation matrix is unknown in practice, and so is the minimum number of subnetworks required for channel autocorrelation matrix recovery by the proposed channel estimation method; whereas applying a larger/smaller number of subnetworks results in higher/lower channel estimation accuracy, and also higher/lower computational complexity. To properly balance the training accuracy and complexity, we propose an efficient progressive training method to train the NN weights. The key idea is to gradually increase the number of subnetworks to generate higher-rank approximations of the exact channel autocorrelation matrix until the training loss converges, without the need for any prior knowledge about the autocorrelation matrix of wideband channel.
  \item Finally, with the estimated channel autocorrelation matrix, the IRS passive reflection coefficients are optimized to maximize the average channel power gain over all OFDM subcarriers. Extensive simulation results are presented to compare the performance of our proposed NN-based IRS channel autocorrelation matrix recovery and passive reflection design with other IRS reflection designs based on user power measurements. Furthermore, we extend the proposed algorithm and validate its effectiveness through comprehensive simulations across diverse system configurations, with different user numbers, IRS array sizes, and power allocation schemes. Simulation results illustrate that our proposed methods can outperform other benchmark schemes under wideband channels with significantly reduced power-measurement overhead and approach the average channel power gain/achievable rate upper bound assuming perfect CSI, thus providing an efficient solution for integrating IRS into OFDM systems.
\end{itemize}

The remainder of this paper is organized as follows. Section \ref{Sec002} presents the IRS-aided OFDM system model. In Section \ref{Sec003}, we derive the expression of user RSRP measurements in terms of the wideband channel autocorrelation matrix and present the proposed single-layer NN structure for its estimation. In Section \ref{Sec004}, the progressive training algorithm is presented to estimate the channel autocorrelation matrix with low complexity, and an IRS reflection design method is presented to optimize the IRS passive reflection subject to the discrete phase-shift constraint based on the estimated wideband channel autocorrelation. Simulation results are presented and analyzed in Section \ref{Sec005}. Section \ref{Sec006} concludes this paper.

\emph{Notations}: $\boldsymbol{I}_M$ stands for the identity matrix of dimension $M$, and $\textbf{0}_{M}$ denotes an all-zero vector of dimension $M$. $\mathbb{R}^{n \times m}$ and $\mathbb{C}^{n \times m}$ stand for the sets of the real-valued and the complex-valued matrices with size of $n \times m$, respectively. The symbol $\cup$ represents the union of two sets, and $\emptyset$ denotes the empty set. $\mathbb{N}$ stands for the set of nonnegative integers. $\left|\cdot\right|$ represents the cardinality of a set or the absolute value of a complex scalar. $\left\|\cdot\right\|$ stands for the Euclidean norm of a vector, and $\left\|\cdot\right\|_F$ denotes the Frobenius norm of a matrix. $\Re\left(\cdot\right)$ and $\Im\left(\cdot\right)$ stand for the real part and the imaginary part of a complex-valued number/vector, respectively. $\jmath = \sqrt{-1}$ denotes the imaginary unit. $\boldsymbol{X}_{n,m}$ represents the element in the $n$-th row and the $m$-th column of a matrix, $\boldsymbol{X}$. ${\cal CN}{\left(\boldsymbol{0}_M,\sigma^2\boldsymbol{I}_M\right)}$ stands for the distribution of a circularly symmetric complex Gaussian (CSCG) random vector that has mean $\boldsymbol{0}_M$ and covariance $\sigma^2\boldsymbol{I}_M$. For a vector/matrix, $\left(\cdot\right)^T$ and $\left(\cdot\right)^H$ represent its transpose and its conjugate transpose, respectively. $\text{rank}(\cdot)$ computes the rank of a matrix, and $\text{Tr}(\cdot)$ returns the trace of a square matrix. $\boldsymbol{R}\succeq0$ indicates that $\boldsymbol{R}$ is a positive semi-definite (PSD) matrix. $\max(\cdot)$ and $\min(\cdot)$ denote the maximum and minimum value of a set of real numbers, respectively. $\text{diag}\left(\boldsymbol{x}\right)$ denotes a square diagonal matrix with $\boldsymbol{x}$ denoting the entries on its main diagonal. $\left\lceil\cdot\right\rceil$ indicates the smallest integer no lower than its argument. $\frac{\partial f}{\partial x}$ stands for the partial derivative of function $f$ with respect to variable $x$. $\nabla_{\boldsymbol x}F \in \mathbb{R}^{n}$ denotes the gradient of a scalar function $F$ with respect to its argument ${\boldsymbol x} \in \mathbb{R}^{n}$. $\mathbb{E}\left[\cdot\right]$ stands for the expectation of a random variable/vector.

\section{System Model}\label{Sec002}

We consider the downlink transmission in an IRS-assisted OFDM system, as depicted in Fig. \ref{fig0001}. The single-antenna BS (or a BS equipped with multiple antennas with fixed downlink precoding) sends downlink messages to a single-antenna user with the assistance of an IRS. A smart controller is linked with the IRS to dynamically adjust the reflection coefficients of its reflecting elements and exchange control messages with a central controller (e.g., the BS or a fusion center) via a separate wireless link. The IRS is endowed with a uniform planar array (UPA) that has $N$ reflecting elements. For the $n$-th IRS reflecting element, its reflection coefficient is denoted by
\begin{equation}\label{Eqs002}
  v_n = \xi_n e^{\jmath \theta_n}, \ n \in {\cal N} \triangleq \left\{1,2, \cdots , N\right\},
\end{equation}
with $\xi_n$ and $\theta_n$ denoting its reflection amplitude and phase shift, respectively. To simplify its hardware design and enhance the reflected signal power from the IRS, we consider $\xi_n=1, n \in {\cal N}$, and only adjust the phase shifts $\theta_n, n \in {\cal N}$ for data transmission. Taking hardware constraints into account, each IRS phase shift can only be configured to a limited number of discrete values, i.e.,
\begin{equation}\label{Eqs002}
  {\theta}_n \in \Phi_\mu \triangleq \{\omega, {2\omega},{3\omega}, \cdots ,2^\mu{\omega}\}, n \in {\cal N},
\end{equation}
where $\mu$ is the number of IRS phase-shift controlling bits, and $\omega = \frac{2\pi}{2^\mu}$.

In this paper, quasi-static block-fading channels that remain approximately unchanged within a channel coherence block are assumed for all links\cite{tse2005fundamentals,goldsmith2005wireless}. This assumption is valid as the IRS is typically deployed at a fixed location and is primarily intended to support low-mobility users in a given target area\cite{Zhengbx2020,Yangyf,Zheng20201} or improve the coverage performance in the target area by maximizing the average power at a number of fixed locations in it that are selected to well represent its spatial power distribution\cite{sunGCC,sunNN1}. In practice, since the IRS is typically located near its served users that are far from the BS for coverage enhancement, the BS-IRS and BS-user wideband channels usually have long propagation distances, thus leading to a large multi-path delay spread. In contrast, the IRS-user wideband channel generally has a smaller multi-path delay spread, due to the close proximity between the IRS and the user\cite{Yangyf,Zheng20201,Zhengsurvey}. For the considered OFDM system, let $K_1$, $K_2$ and $K_3$ denote the maximum number of multi-path delay taps of the BS-user, BS-IRS and IRS-user channels, respectively.
The baseband equivalent channel from the BS to the user, that from the BS to the $n$-th IRS reflecting element and that from the $n$-th IRS reflecting element to the user are denoted by $\boldsymbol{\bar{f}} = \left[ f_{1},f_{2}, \cdots , f_{K_1}\right]^T \in \mathbb{C}^{K_1 \times 1}$, $\boldsymbol{q}_n = \left[ q_{n,1},q_{n,2}, \cdots , q_{n,K_2}\right]^T \in \mathbb{C}^{K_2 \times 1}$ and $\boldsymbol{b}_n = \left[ b_{n,1},b_{n,2}, \cdots , b_{n,K_3}\right]^T \in \mathbb{C}^{K_3 \times 1}$, respectively. Accordingly, the effective time-domain BS-IRS-user cascaded channel via the $n$-th IRS reflecting element can be computed as the convolution of the corresponding BS-IRS channel, IRS-user channel, and the IRS (scalar) reflection coefficient, i.e.,
\begin{equation}
  \boldsymbol{q}_n * v_n * \boldsymbol{b}_n = v_n\boldsymbol{\bar g}_n,
\end{equation}
where $\boldsymbol{\bar g}_n = \boldsymbol{q}_n * \boldsymbol{b}_n \in \mathbb{C}^{K_r \times 1}$ represents the cascaded BS-IRS-user channel (without involving the effect of IRS's phase shifts) via the $n$-th IRS reflecting element, with $K_r = K_2+K_3-1$ denoting the number of delay taps for IRS-cascaded channels.

We assume that $M$ subcarriers have been assigned for the IRS-assisted wideband communication system, where $M \gg \max\left({K_1,K_r}\right) \triangleq K$, with $K$ denoting the maximum number of delay taps for both the cascaded BS-IRS-user channel and the BS-user direct channel (with/without IRS reflection). Let $\boldsymbol{f} = \left[ \boldsymbol{\bar{f}}^T, \textbf{0}_{M-K_1}^T\right]^T \in \mathbb{C}^{M \times 1}$ and $\boldsymbol{g}_n = \left[ {\boldsymbol{\bar g}}_n^T, \textbf{0}_{M-K_r}^T\right]^T \in \mathbb{C}^{M \times 1}$ denote the zero-padded $K_1$-tap BS-user direct channel and $K_r$-tap cascaded BS-IRS-user channel associated with the $n$-th IRS reflecting element, respectively, both of length $M$. Consequently, the superimposed time-domain channel impulse response (CIR) of the wideband channel is expressed as
\begin{equation}
  \tilde{\boldsymbol{h}} = \sum_{n=1}^{N}{v_n\boldsymbol{g}_n} + \boldsymbol{f}.
\end{equation}
By stacking $\boldsymbol{f}$ and $\boldsymbol{g}_n, n \in {\cal N}$, into a matrix $\boldsymbol{G} = \left[\boldsymbol{f}, \boldsymbol{g}_1,\boldsymbol{g}_2,\cdots,\boldsymbol{g}_N\right]\in \mathbb{C}^{M \times \left(N+1\right)}$, the above CIR can be expressed as
\begin{equation}\label{Eqs004}
  \tilde{\boldsymbol{h}} = {\boldsymbol{G}{\boldsymbol{v}}},
\end{equation}
where $\boldsymbol{v} \triangleq \left[1, v_1, v_2, \cdots , v_N\right]^T$ denotes the extended IRS passive reflection vector. Let $\boldsymbol{F}_M$ denote the $M \times M$ discrete Fourier transform (DFT) matrix. The channel frequency response (CFR) of the wideband channel over all of the $M$ subcarriers' thus given by
\begin{equation}\label{Eqs00206}
  \boldsymbol{h} = \boldsymbol{F}_M\tilde{\boldsymbol{h}} = \boldsymbol{F}_M{\boldsymbol{G}{\boldsymbol{v}}}.
\end{equation}

Define $P$ as the BS's transmit power. For convenience, an equal power allocation is applied over the $M$ subcarriers, and the frequency-domain baseband received signal is computed by
\begin{equation}\label{Eqs00207}
  \boldsymbol{y} = \boldsymbol{X}\boldsymbol{h} + \boldsymbol{z},
\end{equation}
where $\boldsymbol{y} \triangleq \left[y_1,y_2, \cdots ,y_M\right]^T \in \mathbb{C}^{M \times 1}$ with $y_m$ denoting the received signal at the $m$-th subcarrier, $\boldsymbol{X}=\text{diag}\left(\boldsymbol{x}\right) \in \mathbb{C}^{M \times M}$ denotes the diagonal matrix of an OFDM symbol $\boldsymbol{x} \in \mathbb{C}^{M \times 1}$ with $\mathbb{E}\left[\left\|\boldsymbol{x}\right\|^2\right]=P$, and $\boldsymbol{z} \triangleq \left[z_1,z_2,\cdots,z_M\right]^T \in \mathbb{C}^{M \times 1} \sim {\cal CN}{\left(\textbf{0}_M,\sigma^2\boldsymbol{I}_M\right)}$ is the receiver noise vector with $\sigma^2$ denoting the average noise power. Accordingly, the average received signal power (ARSP) over all of the $M$ subcarriers is expressed as
\begin{equation}\label{Eqs00208}
  {p}\left(\boldsymbol{v}\right) = \frac{1}{M}\mathbb{E}\left[\left\|\boldsymbol{X}{\boldsymbol{h}} + \boldsymbol{z}\right\|^2\right].
\end{equation}
As $\boldsymbol{z} \sim {\cal CN}{\left(\textbf{0}_M,\sigma^2\boldsymbol{I}_M\right)}$ and is independent of $\boldsymbol{X}$ and $\boldsymbol{h}$, it can be shown that
\begin{equation}\label{Eqs00209}
\begin{split}
  \mathbb{E}\left[\left\|\boldsymbol{X}{\boldsymbol{h}} + \boldsymbol{z}\right\|^2\right] & = \mathbb{E}\left[\left\|\boldsymbol{X}{\boldsymbol{h}}\right\|^2\right] + \mathbb{E}\left[\left\|\boldsymbol{z}\right\|^2\right].
\end{split}
\end{equation}
By substituting (\ref{Eqs00206}) and (\ref{Eqs00209}) into (\ref{Eqs00208}), we have
\begin{align}\label{Eqs002010}
{p}\left(\boldsymbol{v}\right) & = \frac{1}{M}\mathbb{E}\left[\left\|\boldsymbol{X}{\boldsymbol{h}} + \boldsymbol{z}\right\|^2\right] \nonumber \\ & = \frac{1}{M}\mathbb{E}\left[{\boldsymbol{v}^H}{\boldsymbol{G}^H}\boldsymbol{F}_M^H\boldsymbol{X}^H\boldsymbol{X}\boldsymbol{F}_M{\boldsymbol{G}{\boldsymbol{v}}}\right] + \frac{1}{M}\mathbb{E}\left[\left\|\boldsymbol{z}\right\|^2\right]
 \nonumber \\
  & = {\boldsymbol{v}}^H{\boldsymbol{R}}{\boldsymbol{v}} + \sigma^2,
\end{align}
where we have utilized the fact that $\mathbb{E}\left[\boldsymbol{F}_M^H\boldsymbol{X}^H\boldsymbol{X}\boldsymbol{F}_M\right] = P \boldsymbol{I}_M$, and ${\boldsymbol{R}}\triangleq \frac{P}{M}\mathbb{E}\left[{\boldsymbol{G}}^H\boldsymbol{G}\right] \in \mathbb{C}^{(N+1) \times (N+1)}$ denotes the autocorrelation matrix of $\boldsymbol{G}$ (scaled by the power factor $P/{M}$). Note that the channel autocorrelation matrix ${\boldsymbol{R}}$ is Hermitian and PSD (i.e., ${\boldsymbol{R}} \succeq 0$).

In this paper, our objective is to maximize the average channel power gain over all subcarriers by optimizing the IRS's passive reflection subject to the discrete IRS phase-shift constraints. Accordingly, the problem is formulated as
\begin{subequations}\label{Eqs007}
\begin{align}
  (\text{P1}): \ & \max_{\boldsymbol{v}}\frac{{{\boldsymbol{v}}^H{\boldsymbol{R}}{\boldsymbol{v}}}}{P} \label{00701} \\
    & \text{s.t.} \ {\theta}_n \in \Phi_\mu, n \in {\cal N}, \label{00702}
\end{align}
\end{subequations}
where $\boldsymbol{v} \triangleq \left[1, e^{\jmath \theta_1}, e^{\jmath \theta_2}, \cdots , e^{\jmath \theta_N}\right]^T$ denotes the extended IRS passive reflection vector.

There exist various discrete optimization techniques given the perfect knowledge about the channel autocorrelation matrix $\boldsymbol{R}$ to solve problem (P1)\cite{WdIRS,OptAPX}. However, it is difficult to acquire the perfect IRS channel knowledge due to the lack of signal processing ability of IRS as well as the high-dimensional channel parameters in IRS-aided wideband OFDM systems. To tackle this problem, we propose a NN-enabled IRS channel estimation method by utilizing user power measurements for IRS passive reflection design, as detailed next.

\section{NN-enabled Channel Estimation Based on User Power Measurements }\label{Sec003}

In this section, a single-layer NN architecture is proposed to recover the wideband IRS channel autocorrelation matrix based on the user power measurements.

\subsection{RSRP Measurement }

\begin{figure}[t]
  \centering
  {\includegraphics[width=0.35\textwidth]{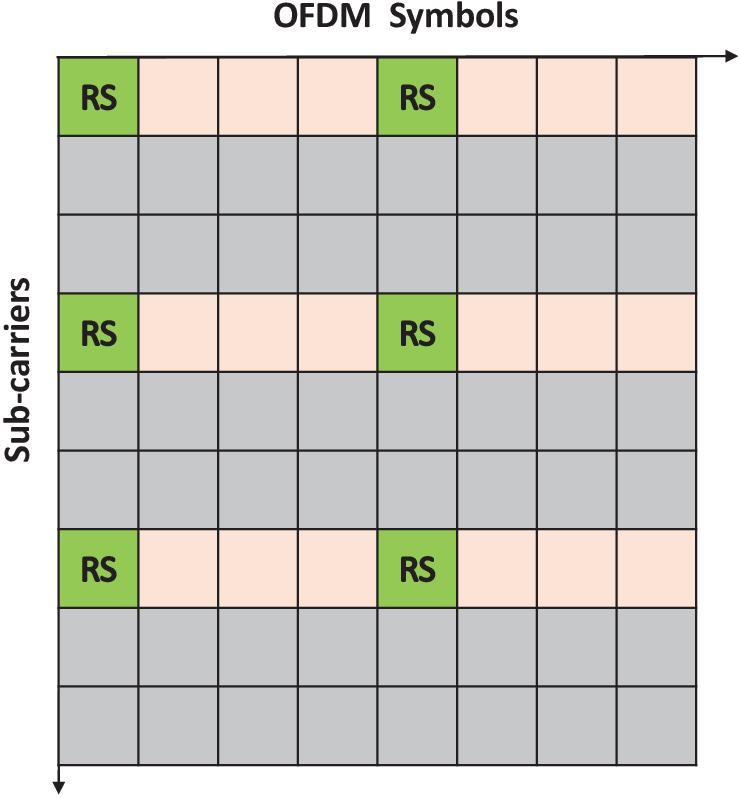}}
  \caption{An example of inserted RSs over OFDM symbols with $Q=2$, $M_0=3$, and $M=9$. }
\label{fig000301}
\end{figure}

In current cellular/WiFi communication systems, the user terminals can measure their received signal power or RSRP. Note that to improve the efficiency of time-frequency resource utilization, reference signals (RSs) are usually inserted in a subset of the $M$ subcarriers for each OFDM symbol\cite{3GPPTS361,3GPPTS36,3GPPTS38_212}, as shown in Fig. \ref{fig000301}. According to the 3GPP standards\cite{3GPPTS361,3GPPTS36}, the downlink reference signal transmit power is equally allocated among the resource elements that carry reference signals which are transmitted by the BS within the system bandwidth. By leveraging the RSRP measurement capability in current cellular/WiFi systems, we propose a practical channel estimation method for the IRS-aided OFDM communication system. Specifically, the IRS is configured with a set of randomly (subject to (\ref{Eqs002})) generated phase shifts for passive reflection training. For each IRS reflection set, the user terminal measures the received power of RSs in $Q$ (non-consecutive) OFDM symbols (see Fig. \ref{fig000301}) in one channel coherence block. Let ${\cal Q} \triangleq \left\{1,2,\cdots,Q\right\}$ denote the set of these OFDM symbols. For each OFDM symbol, we assume that $M_0 (M_0 < M)$ RSs are \emph{uniformly} inserted into $M_0$ subcarriers for power measurement and define ${\cal M}_0 \triangleq \left\{m_1,m_2,\cdots,m_{M_0}\right\}$ as the index set of the subcarriers inserted with RSs.
Based on the above, the RSRP at the user for any given IRS reflection vector $\boldsymbol{v}$ is calculated as
\begin{equation}\label{Eqs00301}
\begin{split}
  \bar{p}\left(\boldsymbol{v}\right) & = \frac{1}{QM_0}\sum_{q \in {\cal{Q}}}{\sum_{m \in {\cal{M}}_0}{\left|{x}_m{{h}_m} + {z}_m(q)\right|^2}},
\end{split}
\end{equation}
where ${z}_m(q)$ denotes the noise at the $m$-th subcarrier in the $q$-th OFDM symbol at the user receiver. In practice, we have $Q \gg 1$ since the symbol rate of OFDM is usually much higher than the reflection switching rate of the IRS. As such, as $Q$ is sufficiently large, the RSRP measurement in (\ref{Eqs00301}) should approach the average received power over RSs in ${\cal M}_0$, i.e.,
\begin{equation}\label{Eqs0030101}
  \bar{p}\left(\boldsymbol{v}\right) \approx \tilde{p}\left(\boldsymbol{v}\right) \triangleq \frac{1}{M_0}\mathbb{E}\left[\left\|\boldsymbol{\bar{X}}{\boldsymbol{\bar{h}}} + \boldsymbol{\bar{z}}\right\|^2\right],
\end{equation}
where $\boldsymbol{\bar{X}}=\text{diag}\left(x_{m_1},x_{m_2},\cdots,x_{m_{M_0}}\right) \in \mathbb{C}^{M_0 \times M_0}$ denotes the diagonal OFDM symbol matrix over the $M_0$ subcarriers in ${\cal M}_0$, $\boldsymbol{\bar{z}} \triangleq \left[z_{m_1},z_{m_2},\cdots,z_{m_{M_0}}\right]^T \in \mathbb{C}^{M_0 \times 1} \sim {\cal CN}{\left(\textbf{0}_{M_0},\sigma^2\boldsymbol{I}_{M_0}\right)}$ denotes the receiver noise vector over the $M_0$ subcarriers, and ${\boldsymbol{\bar{h}}}=\boldsymbol{\bar{F}}\boldsymbol{G}\boldsymbol{v}$ denotes the CFR of the OFDM channel over the $M_0$ subcarriers with $\boldsymbol{\bar{F}} \in \mathbb{C}^{M_0 \times M}$ denoting a partial DFT matrix. Specifically, the $i$-th row of $\boldsymbol{\bar{F}}$ is taken from the $m_i$-th row of $\boldsymbol{F}$, $m_i \in {\cal M}_0$.
Regarding (\ref{Eqs0030101}), we present the following lemma to relate it to (\ref{Eqs002010}).
Note that in long term evolution (LTE) and new radio (NR) systems, the length of cyclic prefix (CP) is typically set to exceed the number of delay taps to mitigate inter-symbol interference (ISI) \cite{hong2022delay}, and the number of RS-inserted subcarriers is usually larger than the CP length\cite{zaidi20185g,3GPPTS38_212} and thus the number of delay taps, i.e., $M_0 \geq K$.

\begin{lemma}\label{lemma0}
Given $M_0 \geq K$, the RSRP measurement in (\ref{Eqs0030101}) based on a subset of subcarriers in ${\cal M}_0$ is equal to the ARSP over all the $M$ subcarriers in (\ref{Eqs002010}), i.e.,
\begin{equation}\label{Eqs00302}
\begin{split}
  {\tilde{p}}\left(\boldsymbol{v}\right) = {p}\left(\boldsymbol{v}\right) = {\boldsymbol{v}}^H{\boldsymbol{R}}{\boldsymbol{v}} + \sigma^2.
\end{split}
\end{equation}
\end{lemma}
The proof of \ref{lemma0} is presented in Appendix \ref{app001}.

Lemma \ref{lemma0} implies that although the RSRP measurement is performed for only a subset of subcarriers, it can characterize the ARSP over all OFDM subcarriers as given in (\ref{Eqs002010}). This suggests that the wideband channel autocorrelation matrix can be extracted from the RSRP measurements.
Let $\boldsymbol{p}=\left[\bar{p}\left(\boldsymbol{v}_1\right),\bar{p}\left(\boldsymbol{v}_2\right),\cdots,\bar{p}\left(\boldsymbol{v}_L\right)\right]$ denote the collection of the user's RSRP measurements under $L$ random IRS training reflection sets with different IRS reflection vectors, $\boldsymbol{v}_1,\boldsymbol{v}_2,\cdots,\boldsymbol{v}_L$. After the above RSRP measurements, the central controller collects them from users for IRS channel estimation.

\subsection{NN-Enabled Channel Autocorrelation Matrix Estimation}

Based on (\ref{Eqs00302}), we aim to recover $\boldsymbol{R}$ based on the RSRP measurements in $\boldsymbol{p}$. However, the channel autocorrelation matrix $\boldsymbol{R} \in \mathbb{C}^{(N+1) \times (N+1)}$ is a high-dimensional Hermitian and PSD matrix, whose estimation entails the complexity in the order of ${\cal O}\left(N^2\right)$. Fortunately, the rank of $\boldsymbol{R}$, which is equal to $K$, is determined by the maximum number of delay taps, where $K \ll (N+1)^2$ in general for IRS with the number of reflection elements $N$. By exploiting this sparsity property of $\boldsymbol{R}$, we can express $\boldsymbol{R}$ equivalently using a small number ($\geq K$) of basis vectors via matrix spectral decomposition, thereby transforming the high-dimensional channel autocorrelation matrix estimation problem into a lower-dimensional basis-vector estimation problem that can be solved with a significantly reduced complexity. To achieve this purpose, we present the following lemma.
\begin{lemma}
\label{lemma1}
For any $N$-dimensional Hermitian matrix $\boldsymbol{R}$ with $\text{rank}\left({\boldsymbol{R}}\right) = K$, there always exist $K'$, $K' \geq K$, basis vectors, $\boldsymbol{a}_k \in \mathbb{C}^{(N+1) \times 1}$, $1 \le k \le K'$, such that
\begin{equation}\label{Eqs00303}
   \boldsymbol{R} = \sum_{k=1}^{K'}{\boldsymbol{a}_k}{\boldsymbol{a}^H_k}, K' \geq K.
\end{equation}
However, if $K' < K$, (15) cannot hold for any $\boldsymbol{a}_k$'s, i.e., we always have
 \begin{equation}\label{Eqs00304}
     \left\|\boldsymbol{R} - \sum_{k=1}^{K'}\boldsymbol{a}_k\boldsymbol{a}^H_k\right\|_F^2>0, \ \forall \boldsymbol{a}_k, K' < K.
   \end{equation}
\end{lemma}
The proof of Lemma \ref{lemma1} is given in Appendix \ref{app002}.

It follows from Lemma \ref{lemma1} that we can always reconstruct $\boldsymbol{R}$ as the sum of $K'(\geq K)$ rank-one matrices, i.e., $\boldsymbol{A}_k \triangleq \boldsymbol{a}_k\boldsymbol{a}_k^H, k=1,2,\cdots,K'$. Based on (\ref{Eqs00303}), if $K' \ge \text{rank}(\boldsymbol{R})$, the noiseless received signal power is given by
\begin{equation}\label{Eqs00308}
  \boldsymbol{v}^H{\boldsymbol{R}}{\boldsymbol{v}}  =\sum_{k=1}^{K'}{{\boldsymbol{v}}^H\boldsymbol{A}_k{\boldsymbol{v}}}=\sum_{k=1}^{K'}{\left|{\boldsymbol{v}}^H\boldsymbol{a}_k\right|^2}.
\end{equation}
Notably, the noiseless received signal power in (\ref{Eqs00308}) can be predicted by a single-layer NN. Particularly, the NN takes the IRS passive reflection vector $\boldsymbol{v}$ and (\ref{Eqs00308}) as its input and output, respectively. Since (\ref{Eqs00308}) is the sum of $K'$ squared amplitude values, i.e., $\lvert \boldsymbol{v}^H \boldsymbol{a}_k \rvert^2, k=1,2,\cdots, K'$, the NN can be divided into $K'$ subnetworks with the weights of the $k$-th subnetwork corresponding to the basis vector $\boldsymbol{a}_k$, as shown in Fig. \ref{fig001}.\footnote{Note that for the special case of $K'=K=1$, this NN reduces to that for the IRS narrowband channel estimation in \cite{sunGCC}, \cite{sunNN1}.} Moreover, to avoid the NN implementation in the complex-value domain\cite{sunGCC}, we can express (\ref{Eqs00308}) in the real-value domain as
\begin{equation}\label{Eqs00309}
  \sum_{k=1}^{K'}{\left|{\boldsymbol{v}}^H\boldsymbol{a}_k\right|^2} = \sum_{k=1}^{K'}{\left\|\boldsymbol{u}^T{\boldsymbol{B}_k}\right\|^2},
\end{equation}
where the complex-valued IRS passive reflection vector $\boldsymbol{v}$ is expressed based on its real part and imaginary part and taken as the NN's input, i.e., $\boldsymbol{u}=\left[{\Re}\left(\boldsymbol{v}^T\right), {\Im}\left(\boldsymbol{v}^T\right)\right]^T \in \mathbb{R}^{\left(2N+2\right) \times 1}$, and $\boldsymbol{B}_k$ is the real-valued basis matrix composed of the real and imaginary parts of $\boldsymbol{{a}}_k$, i.e.,
\begin{equation}\label{Eqs003010}
  \boldsymbol{B}_k = \left[ {\begin{array}{*{20}{c}}
{ {{\mathop{\rm \Re}\nolimits} \left( \boldsymbol{{a}}_k \right)} }&{{{{\mathop{\rm \Im}\nolimits} \left( \boldsymbol{{a}}_k \right)}}}\\
{ {{{\mathop{\rm \Im}\nolimits} \left( \boldsymbol{{a}}_k \right)}}}&{{-{{\mathop{\rm \Re}\nolimits} \left( \boldsymbol{{a}}_k \right)}}}
\end{array}} \right]\in \mathbb{R}^{\left({2N+2}\right) \times 2}.
\end{equation}
Based on (\ref{Eqs00309}), we can establish the single-layer NN structure as shown in Fig. \ref{fig001}, which takes the real-valued IRS passive reflection vector $\boldsymbol{u}$ as its input and aims to recover the noiseless received signal power in (\ref{Eqs00309}) at its output. Particularly, the $K'$ subnetworks share the same input $\boldsymbol{u}$ and aim to recover ${\left\|\boldsymbol{u}^T{\boldsymbol{B}_k}\right\|^2}, k=1,2, \cdots ,K'$, at their individual outputs, respectively. Let $\boldsymbol{W}_{k} \in \mathbb{R}^{\left({2N+2}\right) \times 2}$ denote the weights of the $k$-th subnetwork. As represented in Fig. \ref{fig001}, the weights are labeled by the white squares in each subnetwork. The two neurons in the hidden layer of the $k$-th subnetwork are given by
\begin{equation}\label{Eqs003011}
 \boldsymbol{e}^T_k =\boldsymbol{u}^T{\boldsymbol{W}_k},
\end{equation}
with $\boldsymbol{e}_k \triangleq \left[{e}_{k,1} , {e}_{k,2}\right]^T \in \mathbb{R}^{2}$. The activation function of the $k$-th subnetwork is defined as the squared norm of $\boldsymbol{e}_k$, i.e., ${\left\|\boldsymbol{e}_k\right\|^2}$. Finally, the output of the entire NN is given by the sum of the outputs of the $K'$ subnetworks, i.e.,
\begin{equation}\label{Eqs003012}
  \hat p \left(\boldsymbol{v}\right) = \sum_{k=1}^{K'}{\left\|\boldsymbol{e}_k\right\|^2} = \sum_{k=1}^{K'}{\left\|\boldsymbol{u}^T{\boldsymbol{W}_k}\right\|^2}.
\end{equation}

\begin{figure}[t]
  \centering
  {\includegraphics[width=0.495136\textwidth]{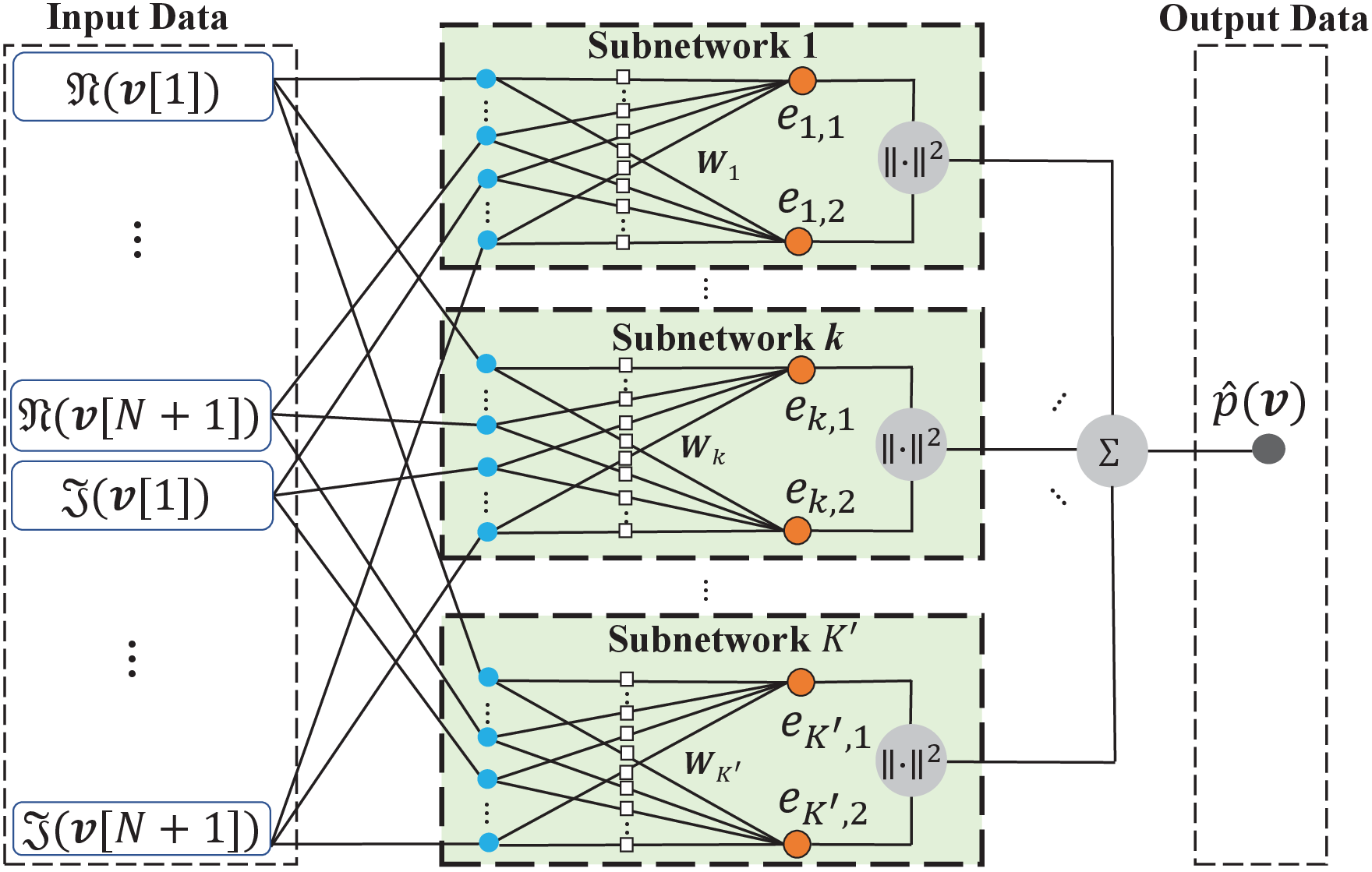}}
  \caption{Single-layer NN structure for wideband channel autocorrelation matrix estimation. }
\label{fig001}
\end{figure}

Comparing (\ref{Eqs00309}) with (\ref{Eqs003012}), it follows that if $\boldsymbol{W}_{k}=\boldsymbol{B}_k, \forall k$, then we have $\hat p\left(\boldsymbol{v}\right) = \boldsymbol{v}^H{\boldsymbol{R}}{\boldsymbol{v}}$. Inspired by this, the real-valued basis matrix $\boldsymbol{B}_k$ can be recovered by training the NN's weight matrix $\boldsymbol{W}_{k}$. To this end, the weight matrix $\boldsymbol{W}_{k}$ of each subnetwork has a similar structure to $\boldsymbol{B}_k$ in (\ref{Eqs003010}), i.e.,
\begin{equation}\label{Eqs001013}
  \boldsymbol{W}_{k} = \left[ {\begin{array}{*{20}{c}}
{ {\boldsymbol{w}_{k,1}} }&{{\boldsymbol{w}_{k,2}}}\\
{ {{\boldsymbol{w}_{k,2}}}}&{{-{\boldsymbol{w}_{k,1}}}}
\end{array}} \right] \in \mathbb{R}^{\left({2N+2}\right) \times 2} ,
\end{equation}
where $\boldsymbol{w}_{k,1} \in \mathbb{R}^{(N+1) \times 1}$ and $\boldsymbol{w}_{k,2} \in {\mathbb{R}}^{(N+1) \times 1}$ correspond to the real and imaginary parts of $\boldsymbol{a}_k$, i.e., ${{{\mathop{\rm \Re}\nolimits} \left( \boldsymbol{{a}}_k \right)}}$ and ${{{{\mathop{\rm \Im}\nolimits} \left( \boldsymbol{{a}}_k \right)}}}$, respectively.

Let $\boldsymbol{w}_{k} = \boldsymbol{w}_{k,1} + \jmath \boldsymbol{w}_{k,2} \in \mathbb{C}^{(N+1) \times 1}$ denote the complex weight vector of the $k$-th subnetwork. Then, we obtain the following lemma.
\begin{lemma}\label{lemma2}
If ${\hat p}\left(\boldsymbol{v}\right) = {\boldsymbol{v}}^H{\boldsymbol{R}}{\boldsymbol{v}}$ holds for any $\boldsymbol{v}$, we have
\begin{equation}\label{Eqs001015}
  \boldsymbol{R} = \sum_{k=1}^{K'}{\boldsymbol{w}_{k}\boldsymbol{w}_{k}^H}.
\end{equation}
\end{lemma}

Proof: See Appendix \ref{app003}.
It is worth noting that Lemma \ref{lemma2} requires that ${\hat p}\left(\boldsymbol{v}\right) = {\boldsymbol{v}}^H{\boldsymbol{R}}{\boldsymbol{v}}$ holds for any $\boldsymbol{v}$, $\boldsymbol{v} \in {\mathbb C}^{N+1}$, while the discrete IRS passive reflection $\boldsymbol{v}$ is restricted to a finite candidate vectors within a subspace of ${\mathbb C}^{N+1}$. Nonetheless, in the case of a large $N$, the size of this subspace for $\boldsymbol{v}$ should be sufficient for (\ref{Eqs001015}) to hold, as will be shown in Section \ref{Sec005}. It follows from the above that we can reconstruct $\boldsymbol{R}$ by training the NN in Fig. \ref{fig001} to estimate $\boldsymbol{W}_k, k = 1, 2, \cdots , K'$, as detailed in the next section. Note that the proposed NN architecture utilizes generic IRS cascaded channels, regardless of specific IRS deployment strategy or the proximity of the IRS to users.

\section{NN Training and IRS Reflection Optimization}\label{Sec004}

In this section, we propose a progressive training method to optimize the NN weights for wideband IRS channel autocorrelation estimation based on user power measurements. Subsequently, the IRS passive reflection is optimized to enhance the average channel power gain for the wideband system based on the estimated IRS channel autocorrelation.

\subsection{Progressive NN Training}

Given the structure of the proposed single-layer NN in Fig. \ref{fig001}, the NN weights can be estimated via supervised learning by employing the user RSRP measurements as training labels. However, the required number of subnetworks, $K'$, depends on the actual channel rank $K$, which is not \emph{a priori} known in practice. To overcome this challenge, we propose a progressive NN training method, which gradually increases the number of subnetworks, $K'$.
Specifically, we start with $K' = \tau$, with $\tau$ denoting the initial number of subnetworks. Then, we gradually increase $K'$ in steps of $\tau$ and sequentially train the corresponding NNs. It is evident that as the number of subnetworks reaches or exceeds the actual channel rank $K$, the lowest approximation error is expected (if ignoring the noise power's effect). As a result, the progressive training can be terminated if the training accuracy cannot be further improved (to be specified later).

\begin{algorithm}[t]
    \caption{Progressive NN Training Algorithm}
    \begin{algorithmic}[1]\label{PNNtrains001}
        \STATE \textbf{Input}: $\boldsymbol{p}$, $\boldsymbol{v}_1,\boldsymbol{v}_2, \cdots ,\boldsymbol{v}_L$, $N$, $L_1$, $\alpha$, $\tau$. \\
        \STATE \textbf{Initialization}: ${K'} = \tau$, $\mathbb{Q} = \emptyset$. \\
        \REPEAT
        \IF {$K'=\tau$}
           \STATE Randomly initialize the weights of NN.
        \ELSE
           \STATE Initialize the weights of the first $K'-\tau$ subnetworks as $\boldsymbol{W}(K'-\tau)$ and randomly initialize the weights of the remaining $\tau$ subnetworks.
        \ENDIF
        \STATE Train the NN by using the SGD method and determine the NN weight matrix $\boldsymbol{W}(K')$ via (\ref{Eqs025}).
        \STATE Compute the minimum MSE based on the validation set, i.e., $\delta(K')$, via (\ref{Eqs031a}).
        \STATE Update $\mathbb{Q}=\mathbb{Q} \cup \{K'\}$ and ${K'} \longleftarrow K' + \tau$.
        \UNTIL $\left| \delta(K')-\delta(K'-\tau) \right| < \varsigma$ or ${K'} > \min(N+1, M)$.
        \STATE Determine the NN weights as $\boldsymbol{W}^\star$ based on (\ref{Eqs032a}).
        \STATE \textbf{Output}: the NN weights $\boldsymbol{W}^\star$.
    \end{algorithmic}
\end{algorithm}

Consider the $n$-th progressive training with $K'=n\tau$, which comprises three stages: weight initialization, weight update, and model validation phases, detailed as follows.

\emph{\underline{Weight Initialization}}: To reduce the NN training error and accelerate training convergence, among the $n\tau$ subnetworks, the weights of the first $(n-1)\tau$ subnetworks can be initialized as those previously trained in the $(n-1)$-th progressive training. As such, we only need to randomly initialize the weights of the newly added $\tau$ subnetworks.

\emph{\underline{Weight Update}}: In the NN training procedure, we randomly select $L_1$ ($L_1<L$) entries of $\boldsymbol{p}$ as the training set, while the remaining $L-L_1$ entries of $\boldsymbol{p}$ are collected in the validation set. Based on Lemma \ref{lemma2}, the loss function is defined as the MSE between the RSRP measurement and the NN's output for the training data, i.e.,
\begin{equation}\label{Eqs019}
  {\cal L} = \frac{1}{L_1}\sum_{l=1}^{L_1}{\left(\bar p(\boldsymbol{v}_l) - \sigma^2 - \hat p(\boldsymbol{v}_l)\right)^2}.
\end{equation}
Given the loss function, the NN weights can be updated using the stochastic gradient descent (SGD) algorithm\cite{SGD}. Based on (\ref{Eqs001013}), the weights of each subnetwork are determined by $2N+2$ entries in $\boldsymbol{W}_k$, i.e., $\boldsymbol{w}_{k,1}$ and $\boldsymbol{w}_{k,2}$. As such, we update only $2N+2$ weights for each subnetwork to ensure that the weight matrix $\boldsymbol{W}_k$ retains the same structure as $\boldsymbol{B}_k$, as shown in (\ref{Eqs001013}), throughout the NN training process. In particular, let $\boldsymbol{W}$ denote the weights of this single-layer NN, which is given by
\begin{equation}\label{Eqs001014}
  \boldsymbol{W} = \left[ \begin{array}{*{20}{c}}
{\boldsymbol{w}_{1,1}}&{\boldsymbol{w}_{2,1}}&{\cdots}&{\boldsymbol{w}_{K',1}}\\
{\boldsymbol{w}_{1,2}}&{\boldsymbol{w}_{2,2}}&{\cdots}&{\boldsymbol{w}_{K',2}}
\end{array}\right] \in {\mathbb{R}}^{\left(2N+2\right) \times K'}.
\end{equation}
Let $\nabla_{\boldsymbol{W}}{{\cal L} }$ denote the gradient of ${{\cal L} }$ with respect to $\boldsymbol{W}$. In the NN training procedure, the weight matrix can be updated as follows:
\begin{equation}\label{Eqs020}
  \boldsymbol{W}^{(t+1)} = \boldsymbol{W}^{(t)} - \alpha\nabla_{\boldsymbol{W}^{(t+1)}}{{\cal L} },
\end{equation}
where $\alpha$ denotes the learning rate and $\boldsymbol{W}^{(t)}$ denotes the value of the weight matrix after the $t$-th iteration. In particular, the gradient $\nabla_{\boldsymbol{W}^{(t+1)}}{{\cal L}_{\boldsymbol{W}}}$ is a matrix with size of $\left(2N + 2\right) \times K'$, where its element in the $j$-th row and the $k$-th column denotes the partial derivative of ${\cal L}_{\boldsymbol{W}}$ with respect to the $j$-th weight in the $k$-th subnetwork with $1 \leq j \leq 2N+2$, $1\leq k \leq K'$, which is given by
\begin{equation}\label{Eqs021}
  \frac{\partial{{\cal L} }}{\partial{W}_{j, k}} = \frac{2}{L_1}\sum_{l=1}^{L_1}\left(\hat p(\boldsymbol{v}_l) - \bar p(\boldsymbol{v}_l) + \sigma^2\right)\frac{\partial{\left\|\boldsymbol{u}^T_l{\boldsymbol{W}_k}\right\|^2}}{\partial{W}_{j, k}},
\end{equation}
where ${W}_{j, k}$ denotes the element in the $j$-th row and the $k$-th column of $\boldsymbol{W}$. The partial derivative in (\ref{Eqs021}) can be computed based on (\ref{Eqs003012}) as
\begin{equation}\label{Eqs022}
  \frac{\partial{\left\|\boldsymbol{u}^T_l{\boldsymbol{W}_k}\right\|^2}}{\partial{{W}}_{j, k}} = 2{e}_{k,1}\frac{\partial{e}_{k,1}}{\partial{{W}}_{j, k}}+2{e}_{k,2}\frac{\partial{e}_{k,2}}{\partial{{W}}_{j, k}},
\end{equation}
with
\begin{equation}\label{Eqs023}
\begin{split}
  \frac{\partial{e}_{k,1}}{\partial{{W}}_{j, k}} & = \left\{ {\begin{array}{*{20}{c}}
{\cos {\theta _{l,j}}},&\text{if}&{1 \le j \le N + 1},\\
{- \sin {\theta _{l,r}}},&\text{if}&{N + 2 \le j \le 2N + 2},
\end{array}} \right. \\
\frac{\partial{e}_{k,2}}{\partial{{W}}_{j, k}} & = \left\{ {\begin{array}{*{20}{c}}
{\sin {\theta _{l,j}}},&\ \ \ \text{if}&{1 \le j \le N + 1},\\
{\cos {\theta _{l,r}}},&\ \ \ \text{if}&{N + 2 \le j \le 2N + 2},
\end{array}} \right.
\end{split}
\end{equation}
where $\theta _{l,r}$ represents the phase shift of the $r$-th IRS reflecting element with its index $r = j - N - 1$.

\emph{\underline{Model Validation}}: After the $t$-th iteration, the MSE between the RSRP measurements and the output of the NN on the validation set is used for model validation, which can be computed by
\begin{equation}\label{Eqs030a}
  \delta^{(t)}(K') = \frac{1}{L-L_1}{\sum_{l=L_1+1}^{L}{\left({ \bar p(\boldsymbol{v}_l) - \sigma^2 - {\hat p}^{(t)}(\boldsymbol{v}_l)}\right)^2}},
\end{equation}
where ${\hat p}^{(t)}(\boldsymbol{v}_l) = \sum_{k=1}^{K'}{\left\|\boldsymbol{u}_l^T{\boldsymbol{W}^{(t)}_{k}}\right\|^2}$ denotes the output of the NN after the $t$-th training iteration, with $\boldsymbol{W}^{(t)}_{k}$ denoting the weight matrix of the $k$-th subnetwork. Denote by $T$ the maximum number of training iterations. As such, the weights of the NN with $K'$ subnetworks are selected as
\begin{equation}\label{Eqs025}
  \boldsymbol{W}(K') = \arg\min_{{1 \leq t \leq T}}{{\delta^{(t)}(K')}},
\end{equation}
and the associated minimum MSE is given by
\begin{equation}\label{Eqs031a}
  \delta(K') = \min_{1\leq t \leq T}{\delta^{(t)}(K')}.
\end{equation}

After the above training process, the progressive training algorithm checks the stopping conditions to determine whether it should proceed to the $(n+1)$-th progressive training with $K'=(n+1)\tau$ or terminate the training process. In particular, the progressive training can be terminated if a desired MSE has been achieved, i.e., $\delta(K')$ is sufficiently small, e.g., when $K' = n\tau \ge K$ is reached. Nonetheless, it is worth noting that the convergence may also be achieved in the case with $K'<K$. In this case, a low-rank approximation to the actual channel autocorrelation matrix can be achieved with a high accuracy, which may occur if the strength of some channel paths is practically low. In addition, as $K \leq \min(N+1, M)$, the training process will also be terminated if $(n+1)\tau \ge \min(N+1, M)$. In summary, the progressive training is terminated if either one of the following two stopping conditions is met:
\begin{enumerate}
  \item The MSE on the validation data set converges, i.e., $\left| \delta(K') - \delta(K'-\tau) \right| < \varsigma$, where $\varsigma$ denotes a prescribed threshold to indicate the convergence of the training error.
  \item The maximum number of subnetworks is reached, i.e., ${K'} +\tau > \min(N+1, M)$.
\end{enumerate}

Based on the above, let $n_{\max}$ denote the number of progressive training required for convergence with $n_{\max} \leq \lceil {\min}(N+1,M)/\tau \rceil$ and $\mathbb{Q} \triangleq \left\{\tau,2\tau,3\tau, \cdots,  n_{\max}\tau\right\}$ denote the set constituting the numbers of subnetworks throughout the progressive training. Next, the NN weights corresponding to the lowest MSE in $\mathbb{Q}$ will be used to recover the channel autocorrelation matrix, i.e.,
\begin{equation}\label{Eqs032a}
  \boldsymbol{W}^\star = \boldsymbol{W}(K^\star), \ \ \ \text{where} \ \ \ K^{\star} = \arg \min_{K' \in {\mathbb{Q}}}{\delta(K')}.
\end{equation}
The procedures of the progressive training algorithm are described in \textbf{Algorithm} \ref{PNNtrains001}. After NN training, the channel autocorrelation matrix can be estimated by
\begin{equation}\label{Eqs033}
  {\boldsymbol{\hat R}} = \sum_{k=1}^{K^{\star}}{\boldsymbol{w}^{\star}_{k}\boldsymbol{w}^{\star H}_{k}},
\end{equation}
where $\boldsymbol{w}^{\star}_{k} = \boldsymbol{w}^{\star}_{k,1} + \jmath \boldsymbol{w}^{\star}_{k,2} \in \mathbb{C}^{\left(N+1\right) \times 1}$ with $\left[\boldsymbol{w}_{k,1}^{{\star}T}, \boldsymbol{w}_{k,2}^{{\star}T}\right]^{T}$ denoting the $k$-th column of $\boldsymbol{W}^\star$.

\subsection{IRS Reflection Design}\label{sec00402}

Based on the estimate in (\ref{Eqs033}), we can optimize the IRS passive reflection by solving problem (P1) via replacing the channel autocorrelation matrix ${\boldsymbol{R}}$ therein with its estimate, ${\boldsymbol{\hat R}}$. However, problem (P1) is a non-convex optimization problem because of the discrete phase-shift constraints. To solve problem (P1) efficiently, we combine the semidefinite relaxation (SDR) technique\cite{SDR} with the successive refinement (SF) strategy\cite{WdIRS}. Define ${\boldsymbol{V}=\boldsymbol{v}\boldsymbol{v}^H}$ as the autocorrelation matrix of reflection vector ${\boldsymbol{v}}$, with $\boldsymbol{V} \succeq 0$. As such, problem (P1) can be reformulated as
\begin{subequations}\label{Eqs00505}
\begin{align}
  \text{(P2):} &\max_{{\boldsymbol{V}}} \ \text{Tr}\left(\boldsymbol{\hat R}\boldsymbol{V}\right) \label{O03016} \\
 & \ \ \text{s.t.} \ {\theta}_n \in \Phi_\mu, \ n=1,2,\cdots,N,\label{C0301604} \\
 & \ \ \ \ \ \ {\boldsymbol{V}}\succeq 0,\label{C0301603} \\
 & \ \ \ \ \ \ \ \text{rank}({\boldsymbol{V}})=1\label{O030165} .
\end{align}
\end{subequations}
However, due to the discrete-phase and rank-one constraints in (\ref{C0301604}) and (\ref{O030165}), problem (P2) remains to be a non-convex optimization problem. To tackle this issue, we relax problem (P2) into a convex semidefinite programming (SDP) problem by replacing the constraint (\ref{C0301604}) with $\boldsymbol{V}_{n,n}=1, n=1, 2, \cdots , N+1$ and removing the constraint (\ref{O030165}). As such, the interior-point method\cite{cvx} can be applied to derive the optimal solution to this SDP problem. Despite that, the achieved solution to the above SDP problem may not satisfy (\ref{C0301604}) or (\ref{O030165}) as required in problem (P2). To address this problem, we perform Gaussian randomization jointly with phase quantization to construct a feasible solution to (P2), denoted by $\widetilde{\boldsymbol{v}}$.

Finally, we apply the SF method to refine the above solution. Specifically, in each refinement iteration, we search for the optimal discrete phase shift for each IRS reflecting element $\theta_n$ to maximize the objective function (\ref{O03016}) in (P2) by enumerating the elements in $\Phi_\mu$, with those of the other entries ${{\theta}_i}, i \ne n$ being fixed. For each IRS reflecting element $\theta_n$, its phase shift is updated as
\begin{equation}
  { \theta }_n^{\star} = \arg\max_{{{ \theta }_n} \in \Phi_\mu } \ \text{Tr}\left(\boldsymbol{\hat R}\boldsymbol{V}\right), \forall n.
\end{equation}
The above refinement iteration is performed until the value of the objective function (\ref{O03016}) converges. The SF method is guaranteed to converge due to the following reasons. First, in each refinement iteration, the optimal solution is achieved for each discrete phase-shift searching, which ensures that the objective value of (P2) is non-decreasing over iterations. Second, the objective value of (P2) is upper-bounded due to the finite BS transmit power.

\begin{remark}
In practice, the wideband channel is predominantly determined by the static scatterers in the environment, such as buildings, ground, and walls, along with the fixed location of the IRS. These dominant channel paths are largely invariant over time, ensuring that the channel autocorrelation matrix ${\boldsymbol{ { R } }}$ remains stable over a long time\cite{LITOFDM,goldsmith2005wireless}. This stability ensures that the optimized IRS reflection pattern, derived from solving (P2) using the estimated channel autocorrelation matrix, can be effectively used for real-time data transmission. Updating the reflection pattern is only required when the dominant channel paths change due to significant environmental variations.
\end{remark}

\begin{remark}
The proposed IRS channel autocorrelation estimation algorithm is also applicable to fast-fading channels to improve average channel power gains over time. This is due to the fact that the estimated IRS channel autocorrelation captures the statistical characteristics of the channel, which remain stable over a large number of channel coherence blocks in fast-fading scenarios, due to the slow-varying scatterer distribution in the propagation environment \cite{tse2005fundamentals}. Utilizing the estimated autocorrelation matrix, the IRS passive reflection can be optimized to improve the average channel power gains in wideband systems.
\end{remark}

Notably, the above RSRP measurements and NN training processes can be performed offline in a target area for coverage improvement, as designed in \cite{sunNN1}. Once the optimized reflection pattern is determined, it can be deployed for real-time data transmission and only requires updates when the statistical channel characteristics experience significant changes. This robustness to small-scale channel variations enables the practical and effective application of the proposed RSRP-based channel autocorrelation estimation method in fast-fading applications.

\section{Numerical Results }\label{Sec005}

In this section, simulation results are presented to evaluate the performance of our proposed IRS channel autocorrelation estimation and the passive reflection design scheme under frequency-selective fading channels. In the three-dimensional Cartesian coordinate system (in meter), as shown in Fig. \ref{fig0001}, an IRS is deployed parallel to the $y$-$z$ plane, which is equipped with a UPA consisting of $N = 4 \times 8 = 32$ reflecting elements with inter-element spacing equal to half wavelength. The reference point (the bottom-left reflecting element) of the IRS is located at $(-2,-1,0)$. The locations of the BS and the receiver are $(35,-20,15)$ and $(0,1,0)$, respectively. Let $\beta_1, \beta_2$ and $\beta_3$ denote the distance-dependent path loss (in dB) of the BS-user, BS-IRS and IRS-user channels, which are given by\cite{YWJSAC,CSM,3GPPTS25996}
\begin{equation}\label{Eqs00501}
  \begin{split}
    \beta_1 & =  33 + 37\text{log}_{10}(d_1), \\
    \beta_2 & =  30 + 20\text{log}_{10}(d_2), \\
    \beta_3 & =  30 + 20\text{log}_{10}(d_3),
  \end{split}
\end{equation}
with $d_1,d_2$ and $d_3$ denoting the distance in meter between the BS and the user, that between the BS and the IRS, and that between the IRS and the user, respectively.

The OFDM system is configured with $M = 128$ subcarriers, with the number of delay taps for the BS-user, BS-IRS and IRS-user channels set to $K_1=4$, $K_2=4$ and $K_3=3$, respectively. Considering that the BS is far away from the IRS/user, the BS-IRS and BS-user frequency-selective channels are modeled using an exponentially decaying power delay profile given by $\zeta_{i , k }=\frac{1}{\eta_{i , k}}e^{-\varepsilon(k-1)}$, $k=1,2, \cdots , K_i, i=1,2$, where $\eta_{i,k}=\sum_{k=1}^{K_i}{e^{-\varepsilon(k-1)}}$ denotes the normalization factor with $\varepsilon$ denoting the exponential decay factor\cite{Powerechannel0,zhu2024performance}. The CIR coefficients for each link are generated based on independent and identically distributed (i.i.d.) Rayleigh fading with average power $10^{-\beta_i/10}\zeta_{i,k}, k=1,2, \cdots , K_i, i=1,2$\cite{Zhengbx2020}. Besides, we assume frequency-selective Rician fading for the IRS-user link. Specifically, the first tap of the IRS-user link is modeled as the LoS component, while the other taps are modeled as i.i.d. NLoS Rayleigh fading components. Let $\kappa$ denote the Rician factor, i.e., the power ratio of the dominant LoS component to that of all the NLoS fading components. In the simulation, we set $\kappa=7$ dB\cite{3GPPTS25996}. The number of RS-inserted OFDM symbols is set to $Q=30$, and the number of RSs inserted in each OFDM symbol is set to $M_0=64$\cite{3GPPTS36,3GPPTS38_212}. The noise power is $\sigma^2=-90$ \text{dBm} and the BS's transmit power is $P=30$ \text{dBm}. The number of controlling bits for IRS reflection phase shifts is set to $\mu = 2$. The power decaying factor in the OFDM channel is $\varepsilon=2$, if not specified otherwise.

\begin{figure}[t]\hspace{0.336cm}
  {\includegraphics[width=0.5\textwidth]{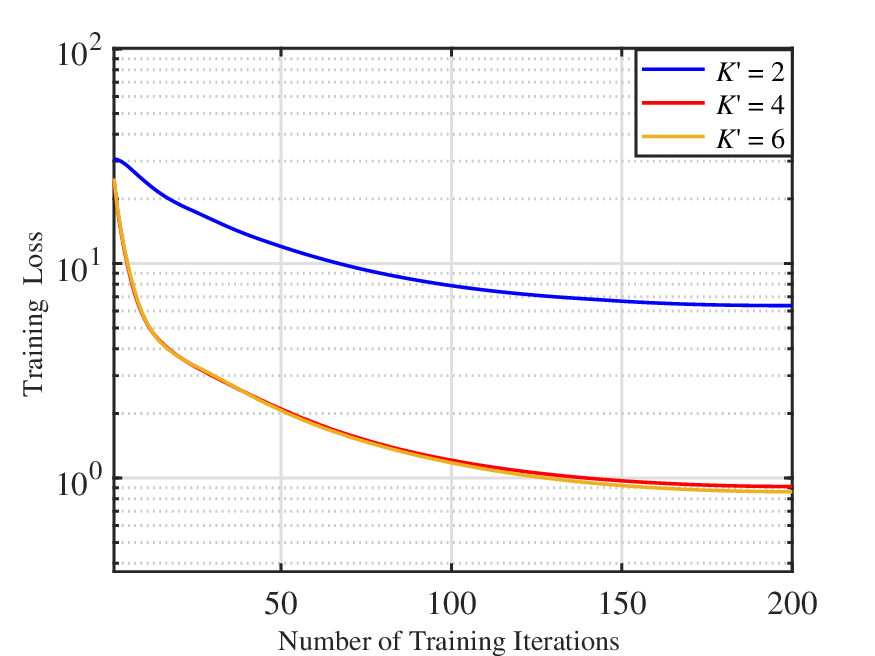}}
  \caption{NN training loss versus the number of training iterations. }
\label{fig00501}
\end{figure}

\begin{figure}[t]\hspace{0.336cm}
  {\includegraphics[width=0.5\textwidth]{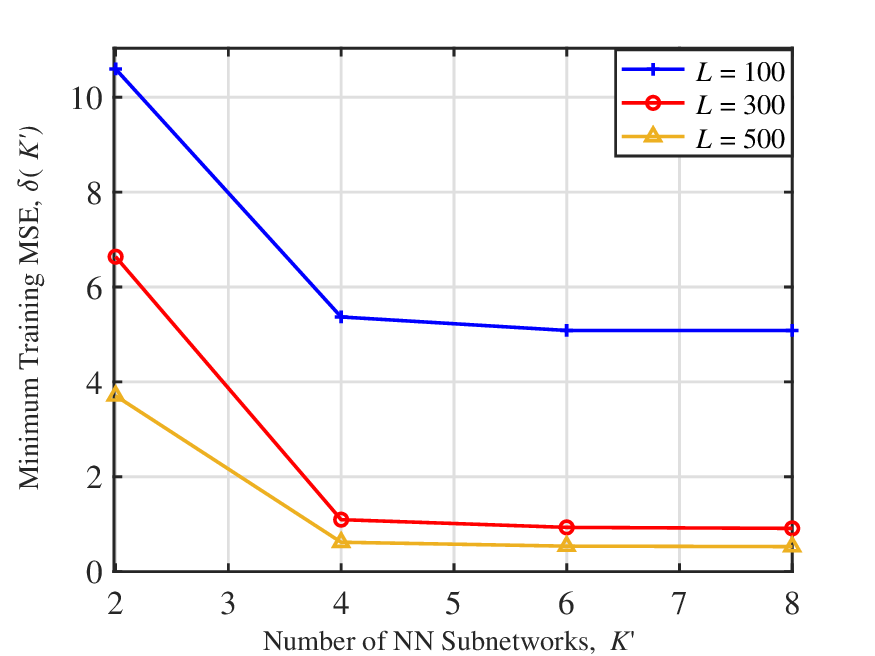}}
  \caption{Minimum training MSE of progressive training versus the number of NN subnetworks, $K'$. }
\label{fig00502}
\end{figure}

\begin{figure}[t]
\hspace{0.336cm}
  {\includegraphics[width=0.5\textwidth]{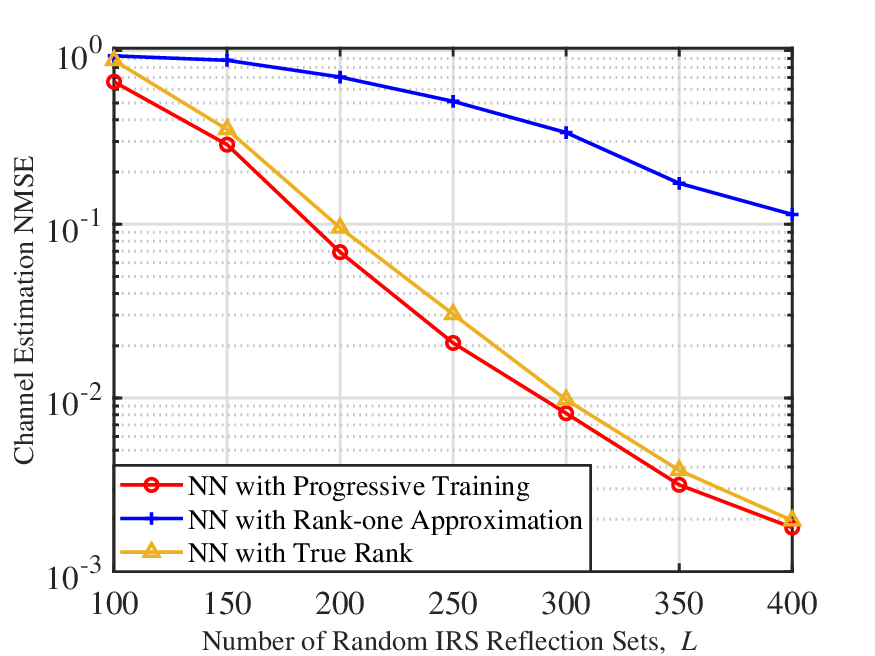}}
  \caption{NMSE of IRS channel autocorrelation estimation versus $L$. }
\label{fig0401}
\end{figure}

\subsection{Progressive NN Training}

In the progressive training algorithm (\textbf{Algorithm} \ref{PNNtrains001}), the NN is trained via the SGD optimizer \cite{SGD} with an initial learning rate $\alpha=10^{-3}$, and the learning rate drops by $0.5\%$ at each training iteration. The threshold for identifying the convergence of the training error is set to $\varsigma=0.1$. For each NN with a given number of subnetworks, the number of training iterations is set to $T=200$. We evaluate the convergence of the NN-based channel estimation algorithm by examining its training loss function values during the progressive training. Fig. \ref{fig00501} plots the training loss versus the number of NN training iterations with $L=300$ random IRS reflection sets. It is observed that the training loss decreases rapidly with the number of NN training iterations. Fig. \ref{fig00502} shows the minimum training MSE $\delta(K')$ over the validation data set versus the number of subnetworks. It depicts that as $K'<4$, the minimum MSE drops rapidly with increasing $K'$. However, as $K'>4$, increasing $K'$ has little effect on reducing the minimum MSE, which indicates that a sufficient number of subnetworks (i.e., $K'=4$) has already been achieved. In addition, Fig. \ref{fig00502} shows that the minimum MSE decreases with increasing $L$, thanks to the enlarged training set.

\subsection{IRS Channel Autocorrelation Matrix Estimation}\label{sec0501}

Next, the normalized MSE (NMSE) performance of the progressive NN training algorithm for IRS channel autocorrelation matrix estimation is evaluated over $1000$ independent channel realizations. In particular, the NMSE between the estimated channel autocorrelation matrix and the actual one is given by
\begin{equation}\label{Eqs00502}
  \text{NMSE} = \frac{\left\|\boldsymbol{\hat R} - \boldsymbol{R}\right\|_F^2}{\left\|\boldsymbol{R}\right\|_F^2}.
\end{equation}

Moreover, we show the NMSE by the proposed NN-enabled channel estimation with the number of subnetworks $K'=1$ (labeled as ``NN with Rank-one Approximation'') or $K'=K$ (labeled as ``NN with True Rank'') as two benchmarks to evaluate the effectiveness of the progressive training algorithm. Note that by setting $K'=1$, the proposed single-layer NN reduces to that for estimating the narrowband rank-one IRS channel autocorrelation matrix \cite{sunGCC}, \cite{sunNN1}.

Fig. \ref{fig0401} shows the NMSE of IRS channel estimation versus the number of random IRS reflection sets, $L$. It depicts that the NMSE decreases by enlarging $L$, as expected. In addition, the proposed progressive NN training algorithm achieves much lower NMSE than the benchmark of Rank-one Approximation, owing to its ability to generate a higher-rank approximation of the channel autocorrelation matrix under the wideband setup. Moreover, it is interesting to observe that the progressive NN training can even outperform the benchmark of ``NN with True Rank'' when the set of training data is small. This is because the progressive training enables adaptive tuning of the NN's size and the total count of training iterations based on the noisy training data available.

\begin{figure}[t]
\hspace{0.336cm}
  {\includegraphics[width=0.5\textwidth]{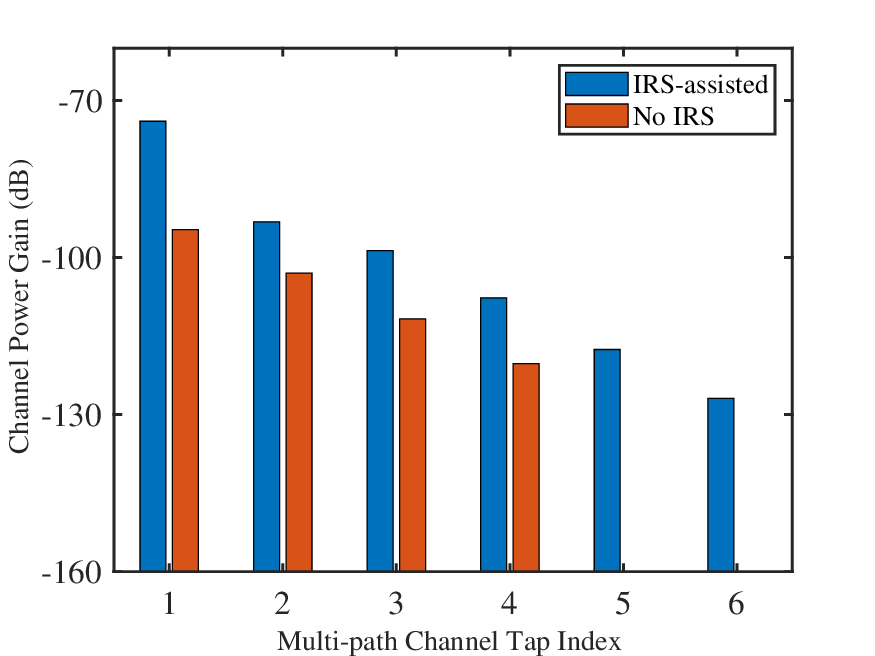}}
  \caption{Power gains over multi-path channel taps with versus without IRS. }
\label{fig00505}
\end{figure}

\subsection{IRS Passive Reflection Design}

To demonstrate the effectiveness of IRS passive reflection optimization utilizing the proposed channel autocorrelation matrix estimation, we first show in Fig. \ref{fig00505} the average power gain of each multi-path channel tap with versus without IRS over $1000$ independent channel realizations. In the case with IRS, its passive reflection is optimized by solving problem (P2) based on the channel autocorrelation matrix estimated by our proposed NN-based algorithm with $L=350$ RSRP measurements. It illustrates that the power gain of each multi-path channel tap is significantly improved after IRS passive reflection optimization. Moreover, there are $K_r-K_1=2$ additional taps after deploying the IRS, which helps improve the average channel power gain in the OFDM communication system.

Next, we compare the average channel power gain at the user in (\ref{00701}) by different IRS passive reflection designs. In particular, we adopt the random-max sampling (RMS)\cite{CSM,tao2020intelligent} and CSM \cite{CSM} methods as benchmark schemes, as both of them utilize user power measurements for optimizing the IRS passive reflection. Specifically, the RMS method selects the IRS reflection set that yields the maximum received signal power among all RSRP measurements as the optimized one, which is given by
\begin{equation}\label{s3001}
  {\boldsymbol{v}}^{\text{RMS}} = \boldsymbol{v}_{l^{\star}}, \ \ \ \text{with} \ \ \ {l^{\star}}=\arg{{\max_{1 \leq l \leq L}}}{\bar{p}}(\boldsymbol{v}_l) .
\end{equation}
While the CSM method in \cite{CSM} computes the sample mean of RSRP measurements conditioned on $\theta_n=\psi, \psi \in \Phi_\mu$, i.e.,
\begin{equation}\label{s3002}
  {{\mathbb{E}}}[p|\theta_n=\psi] = \frac{1}{\left| \mathcal{S}_{n}(\psi) \right|} \sum_{\boldsymbol{v} \in \mathcal{S}_{n}(\psi)}{{{\bar{p}}(\boldsymbol{v})}},
\end{equation}
where $\mathcal{S}_{n}(\psi)$ denotes the subset of the $L$ random IRS passive reflection sets restricted to $\theta_n=\psi$, $\forall n$. Accordingly, the $n$-th IRS phase shift is determined by
\begin{equation}
\label{s3003}
  \theta_n^{\text{CSM}} = \arg{\max_{\psi \in \Phi_\mu}{{{\mathbb{E}}}[p|\theta_n=\psi]}}, \ n = 1, 2,  \cdots ,N.
\end{equation}

Furthermore, the IRS reflection design using perfect CSI (i.e., by substituting the perfect knowledge of ${\boldsymbol{R}}$ into (\ref{O03016}) for solving problem (P2)) is also included as the performance upper bound on the achievable average channel power gain (labeled as ``Upper Bound'').

\begin{figure*}[t]
  \centering
  \subfigure[$P=25$ \text{dBm}, $\mu=1$ ]{\includegraphics[width=0.45\textwidth]{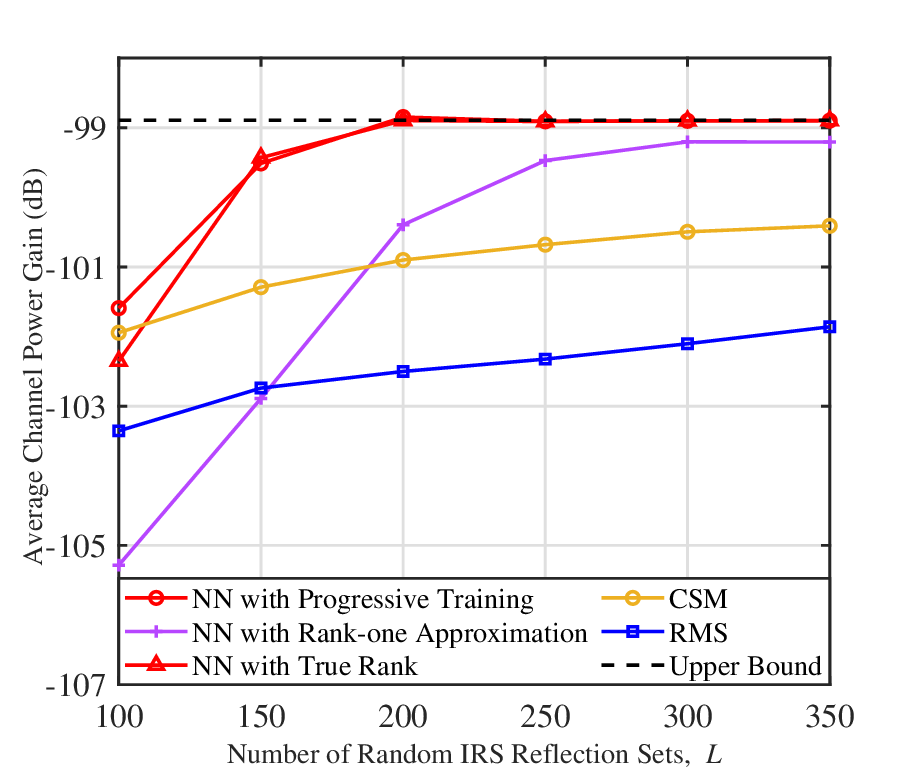}}\hspace{0.31336cm}
  \subfigure[$P=30$ \text{dBm}, $\mu=1$ ]{\includegraphics[width=0.45\textwidth]{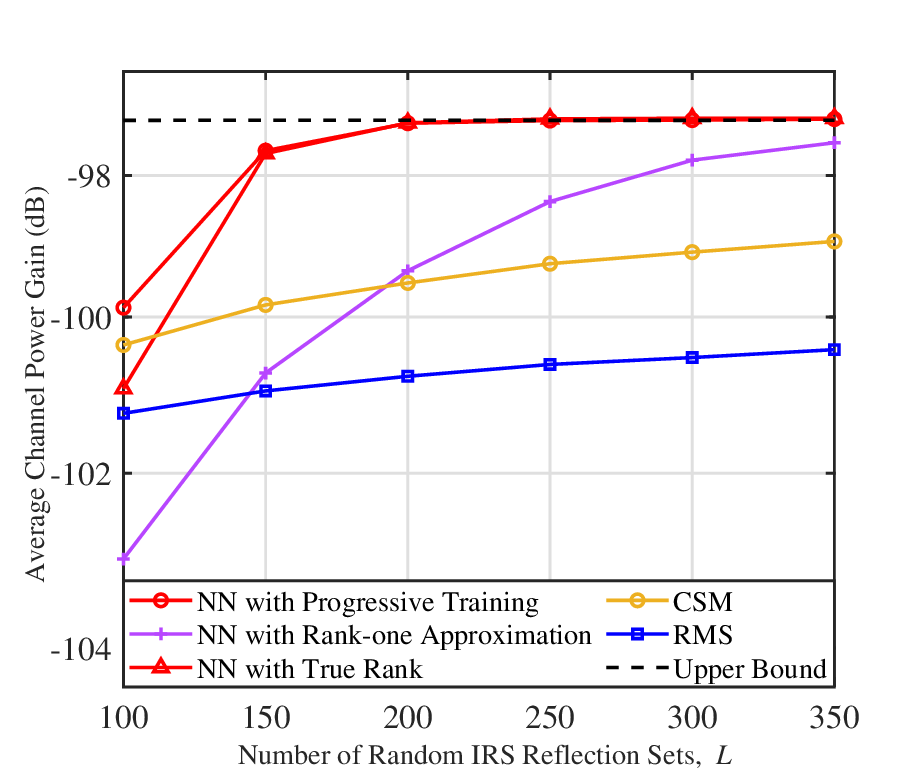}}
  \subfigure[$P=25$ \text{dBm}, $\mu=2$ ]{\includegraphics[width=0.45\textwidth]{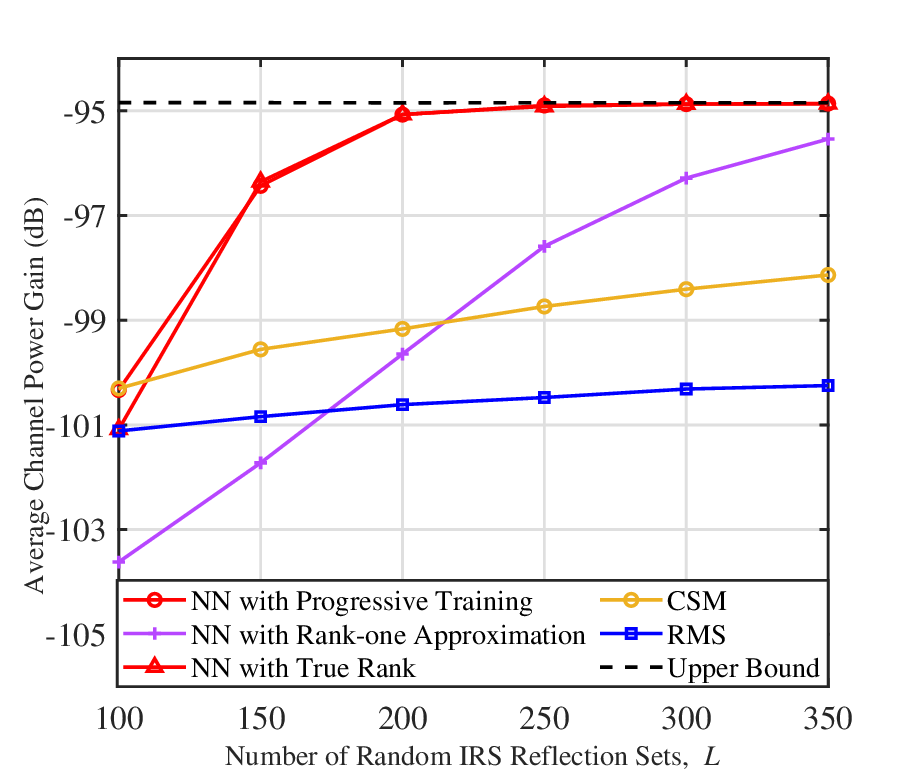}}\hspace{0.6cm}
  \subfigure[$P=30$ \text{dBm}, $\mu=2$ ]{\includegraphics[width=0.45\textwidth]{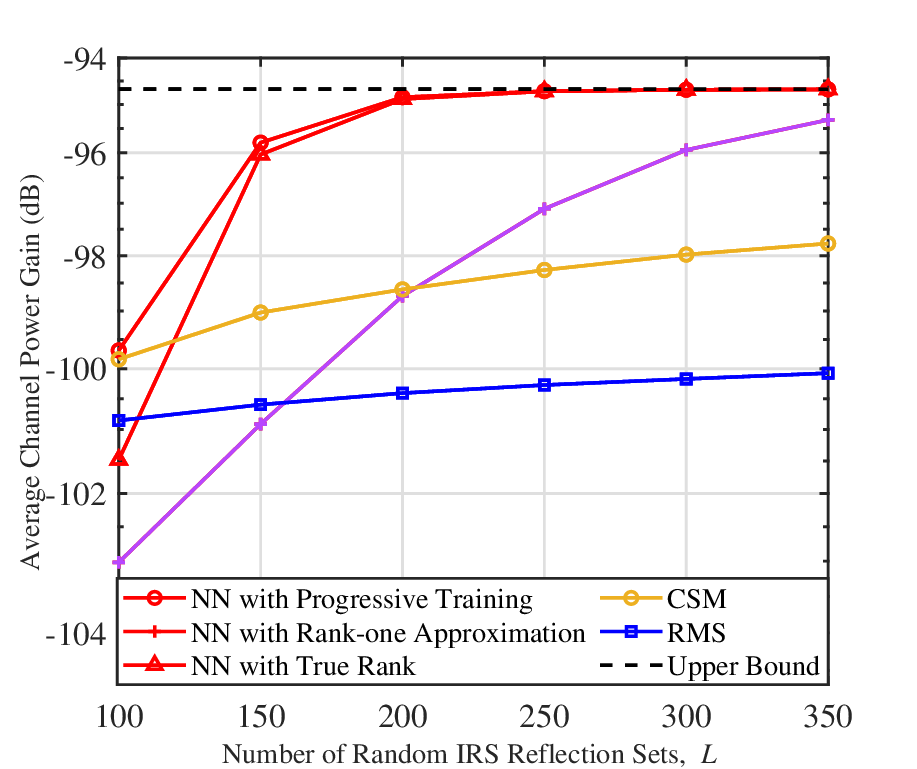}}
  \caption{Average channel power gain versus the number of random IRS reflection sets, $L$. }
\label{fig0005}
\vspace{-9.61pt}
\end{figure*}

In Figs. \ref{fig0005}(a) and \ref{fig0005}(b), we plot the average channel power gain achieved by different IRS passive reflection designs versus the number of random IRS passive reflections, $L$, under $\mu = 1$, with $P=25$ dBm and $P=30$ dBm, respectively. The results depict that the average channel power gain achieved by the proposed progressive NN training method is higher than that by other benchmark schemes. Additionally, although a large performance gap exists between the ``NN with Rank-one Approximation'' and the ``NN with True Rank'', the former can still outperform both the CSM and RMS benchmarks given a sufficiently large number of random IRS reflection sets, owing to its capability of rank-one channel autocorrelation matrix approximation. In addition, the progressive NN training scheme achieves better average channel power gain performance than ``NN with True Rank'' when $L$ is small, as similarly observed in Fig. \ref{fig0401}.

Moreover, the proposed progressive NN training scheme achieves a larger improvement in average channel power gain compared to ``NN with Rank-one Approximation'', due to the high-rank matrix approximation of the channel autocorrelation matrix based on multiple subnetworks. This observation indicates that compared to using a single subnetwork, using multiple subnetworks can more accurately recover the channel autocorrelation matrix, thereby achieving higher average channel power gains via IRS reflection optimization in wideband systems. Furthermore, by further increasing $L$, its achieved average channel power gain can approach the upper bound assuming perfect CSI.

Figs. \ref{fig0005}(c) and \ref{fig0005}(d) show the average channel power gain under different number of random IRS passive reflections $L$ with the number of controlling bits $\mu = 2$. Similar observations to Figs. \ref{fig0005}(a) and \ref{fig0005}(b) can be made in Figs. \ref{fig0005}(c) and \ref{fig0005}(d). In addition, a significant average channel power gain improvement by the proposed scheme is observed after increasing $\mu$, owing to the improved resolution of IRS phase shift for both IRS channel autocorrelation matrix recovery and IRS reflection optimization.

Fig. \ref{fig0a9} shows the average channel power gain versus the number of random IRS passive reflections $L$ with the number of IRS reflection elements $N= 8 \times 16 = 128$, the number of controlling bits $\mu = 2$ and $P=30$ dBm. Similar observations for $N=32$ in Fig. \ref{fig0005}(d) can also be made in Fig. \ref{fig0a9}. As compared to Fig. \ref{fig0005}(d), the maximum average channel power gain is further improved by increasing the number of reflecting elements, due to the increased array gain and beamforming gain provided by the IRS.
\vspace{-16pt}

\begin{figure}[t]
  \centering
  {\includegraphics[width=0.5\textwidth]{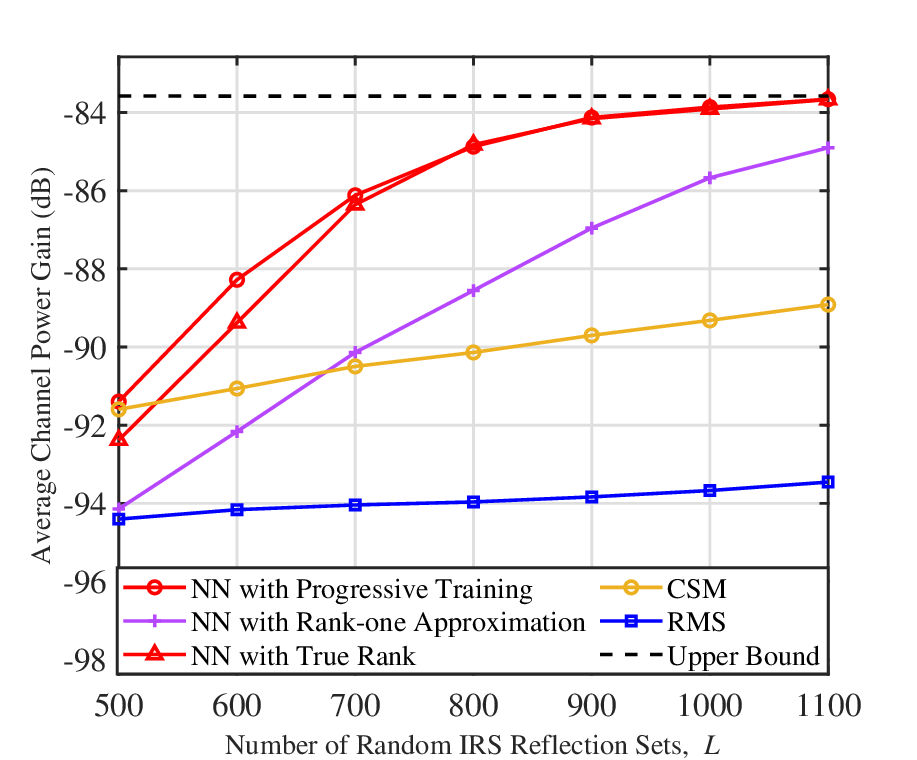}} 
  \caption{Average channel power gain with an IRS array of size $N=128$. }
\label{fig0a9}
\vspace{-9.16pt}
\end{figure}

\begin{figure}[t]
  \centering
  {\includegraphics[width=0.5\textwidth]{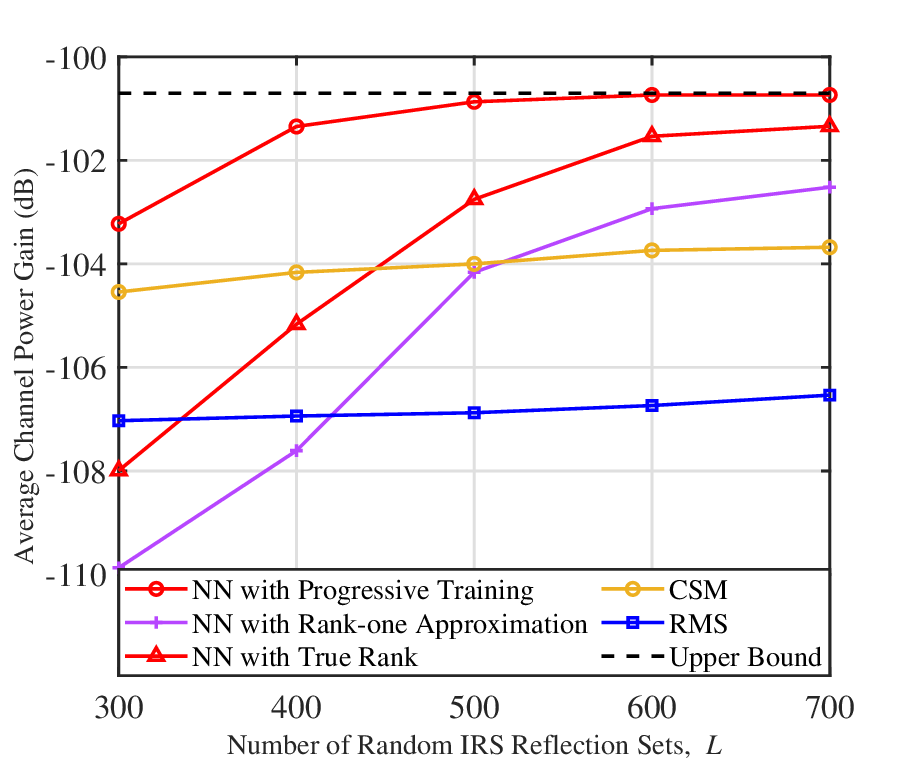}} 
  \caption{Average channel power gain in multiuser scenarios. }
\label{fig0a10}
\end{figure}

\subsection{Extension to Multi-user Systems}
The proposed IRS channel autocorrelation estimation and reflection design method can be extended to multiuser systems. In particular, we consider a scenario with five users whose locations are randomly distributed within a square area defined by the corner coordinates $(0,0,0), (0,10,0), (10,10,0)$ and $(10,0,0)$. An IRS with $N= 8 \times 8$ reflecting elements is deployed to serve the users. In the multiuser scenario, the IRS channel autocorrelation estimation method can be implemented by independently training a NN for each user to reconstruct their respective IRS channel autocorrelation matrices\cite{sunGCC}. Since the proposed IRS channel autocorrelation estimation method relies only on the downlink received signal power measurements under random IRS training reflection patterns, the set of IRS training reflection patterns (i.e., $\boldsymbol{v}_1, \boldsymbol{v}_2, \cdots, \boldsymbol{v}_L$) used for RSRP measurements can be shared by all users. Based on the estimated IRS channel autocorrelation matrices, the IRS reflection is optimized by solving (P2) with $\boldsymbol{\hat R}$ replaced by the average of the estimated channel autocorrelation matrices of all users\footnote{The estimated channel autocorrelation matrices of all users can also be utilized to maximize the minimum channel power gain among all users. The detailed procedures for solving this problem can be found in \cite{sunGCC} and thus are omitted here for brevity.}. Fig. \ref{fig0a10} shows the average channel power gain of all users. It can be observed that the proposed scheme significantly outperforms the benchmark schemes by exploiting the power measurements in multiuser scenarios.

\vspace{9.16pt}
\subsection{Effect of Channel Power Decaying Factor}

\begin{figure}[t]
  \centering
  {\includegraphics[width=0.5\textwidth]{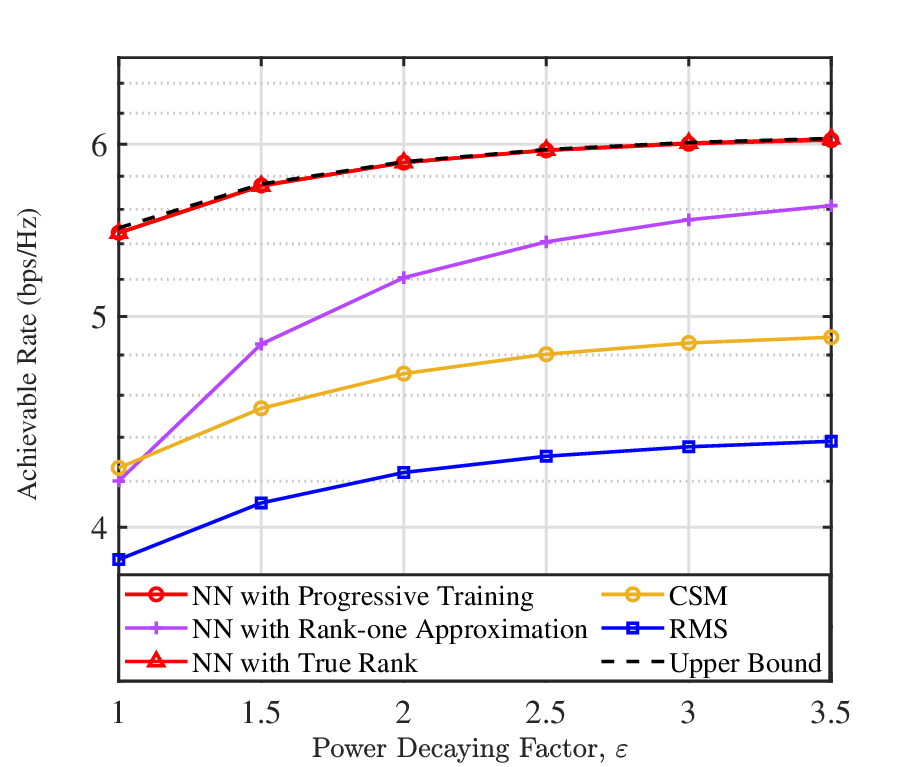}}
  \caption{OFDM achievable rate versus power decaying factor, $\varepsilon$. }
\label{fig07006}
\end{figure}

Finally, we investigate the achievable rate of the OFDM system under different channel power decaying factors. The OFDM achievable rate in bits per second per Hertz (bps/Hz) is given by
\begin{equation}
   \mathcal{C}= \frac{1}{M+M_{\text{CP}}}\sum_{m=1}^M{\text{log}_2\left(1+\frac{P_m\left|h_m^\star\right|^2}{\sigma^2}\right)},
\end{equation}
where $M_{\text{CP}}$ denotes the length of cyclic prefix, and $h_m^\star$ denotes the effective BS-user channel over the $m$-th subcarrier, i.e., the $m$-th entry of the CFR vector $\boldsymbol{F}_M{\boldsymbol{G}{\boldsymbol{v}^\star}}$ with $\boldsymbol{v}^\star$ denoting the optimized IRS passive reflection. $P_m$ denotes the transmit power allocated to the $m$-th subcarrier, and the optimal transmit power allocation can be determined by the water-filling algorithm\cite{LITOFDM,zhu2024performance}.
Fig. \ref{fig07006} depicts the achievable rates by different IRS reflection design schemes under different channel power decaying factors, $\varepsilon$, with the number of IRS phase-shift controlling bits $\mu = 2$ and the CP length $M_{\text{CP}}=16$. The number of random IRS reflection sets is configured to $L=300$, and the BS's transmit power is $P=25$ dBm. First, it is observed that our proposed scheme yields a higher OFDM achievable rate than those of other benchmarks and approaches the rate upper bound with perfect CSI in the scenario with optimal power allocation among subcarriers.
Second, it illustrates that the OFDM achievable rate improves as the decaying factor increases. This is because with a large decaying factor, the power of the received signal will be concentrated on less number of taps. Consequently, the IRS passive reflection design can achieve a higher passive beamforming gain by aligning its beams with dominant paths, thereby improving the rate performance. Third, the performance gap between ``NN with Rank-one Approximation'' and the rate upper bound decreases by increasing the decaying factor $\varepsilon$. This implies that in the case of a large decaying factor, it suffices to train the NN with a single subnetwork similar to that for the narrowband system in \cite{sunGCC}, \cite{sunNN1} with low performance loss, thus reducing the training complexity.

\section{Conclusion}\label{Sec006}

In this paper, a single-layer NN-facilitated wideband IRS channel autocorrelation estimation and passive reflection design framework based on user power/RSRP measurements was proposed for OFDM communication systems. First,we derived the user's average received power over all OFDM subcarriers in terms of the IRS wideband channel autocorrelation matrix, showing that it could be efficiently recovered by optimizing the weights of a single-layer NN consisting of multiple subnetworks based on the user's RSRP measurements over only a subset of subcarriers. To reduce the training complexity without knowing the actual rank of the channel autocorrelation matrix, we proposed a progressive NN training method by gradually increasing the number of subnetworks until convergence is reached. Simulation results demonstrated that the proposed IRS channel autocorrelation matrix estimation and reflection optimization framework substantially surpasses other power-measurement-based benchmarks and approaches the performance upper bound assuming perfect CSI under the wideband channel setup. Additionally, extensive simulations validated the effectiveness of the proposed method in various system setups. Our proposed wideband IRS channel autocorrelation matrix estimation design can be extended to multi-IRS communication systems, which are worth studying in future work.

\appendix

\subsection{Proof of Lemma \ref{lemma0}}\label{app001}
By taking advantage of the fact that $\mathbb{E}\left[\boldsymbol{\bar{X}}{\boldsymbol{\bar{h}}}\boldsymbol{\bar{z}}\right]=0$, (\ref{Eqs0030101}) can be simplified as
\begin{equation}\label{Eqs0A001}
\tilde{p}\left(\boldsymbol{v}\right) = \frac{1}{M_0}\mathbb{E}\left[\left\|\boldsymbol{\bar{X}}{\boldsymbol{\bar{h}}}\right\|^2\right] + \sigma^2.
\end{equation}
By substituting ${\boldsymbol{\bar{h}}}=\boldsymbol{\bar{F}}\boldsymbol{G}\boldsymbol{v}$ into (\ref{Eqs0A001}), we have
\begin{equation}\label{Eqs0A002}
\begin{split}
  \tilde{p}\left(\boldsymbol{v}\right) & = \frac{1}{M_0}\mathbb{E}\left[\left\|\boldsymbol{\bar{X}}{\boldsymbol{\bar{F}}\boldsymbol{G}\boldsymbol{v}}\right\|^2\right] + \sigma^2 \\
    & = \frac{P}{M_0M}{\boldsymbol{v}^H\mathbb{E}\left[\boldsymbol{G}^H\boldsymbol{\bar{F}}^H\boldsymbol{\bar{F}}\boldsymbol{G}\right]\boldsymbol{v}} + \sigma^2 ,
\end{split}
\end{equation}
where we have utilized the fact that $\mathbb{E}\left[\boldsymbol{\bar{X}}^H\boldsymbol{\bar{X}}\right] = \frac{P}{M}\boldsymbol{I}_{M_0}$. It is worth noting that the partial DFT matrix $\boldsymbol{\bar{F}}$ is constructed by selecting the $m_i$-th row ($\forall \ m_{i} \in {{\cal M}_0}$) from the DFT matrix $\boldsymbol{F}_{M}$. As the RSs are uniformly inserted into $M_0$ subcarriers, we have $m_{i+1}-m_{i}=m_{i+2}-m_{i+1}, i =1,2,\cdots,M_0-2, \forall m_{i} \in {{\cal M}_0}$. Thus, the autocorrelation $\boldsymbol{\bar{F}}^H\boldsymbol{\bar{F}}$ of the partial DFT matrix is cyclic symmetric and given by\footnote{Note that the total number of subcarriers is divisible by the number of subcarriers inserted with RSs, i.e., $M/M_0 \in \mathbb{N}$, as specified by the 3GPP standard\cite{3GPPTS38_212}.}
\begin{equation}\label{Eqs0A003}
\boldsymbol{\bar{F}}^H\boldsymbol{\bar{F}} = M_0\left[ {\begin{array}{*{20}{c}}
{{\boldsymbol{I}_{{M_0}}}}&{{\boldsymbol{I}_{{M_0}}}}& \cdots &{{\boldsymbol{I}_{{M_0}}}}\\
{{\boldsymbol{I}_{{M_0}}}}&{{\boldsymbol{I}_{{M_0}}}}& \cdots &{{\boldsymbol{I}_{{M_0}}}}\\
 \vdots & \vdots & \ddots & \vdots \\
{{\boldsymbol{I}_{{M_0}}}}&{{\boldsymbol{I}_{{M_0}}}}& \cdots &{{\boldsymbol{I}_{{M_0}}}}
\end{array}} \right] \in \mathbb{C}^{M \times M}.
\end{equation}
Furthermore, note that the last $(M-K)$ rows of the CIR matrix $\boldsymbol{G}$ are zero-padded, i.e.,
\begin{equation}\label{Eqs0A004}
  \boldsymbol{G} = \left[ {\begin{array}{*{20}{c}}
\boldsymbol{\bar{G}}\\
\textbf{0}_{(M-K) \times (N+1)}
\end{array}} \right] \in \mathbb{C}^{M \times (N+1)},
\end{equation}
where $\boldsymbol{\bar{G}} \in \mathbb{C}^{K \times (N+1)}$ denotes the first $K$ rows of the CIR matrix, $\boldsymbol{G}$.
Thus, we have
\begin{equation}\label{Eqs0A00502}
\boldsymbol{\bar {G}}^H\boldsymbol{\bar {G}} = \boldsymbol{G}^H\boldsymbol{G}.
\end{equation}
Since $M_0 \geq K$, based on (\ref{Eqs0A003}) and (\ref{Eqs0A004}), we have
\begin{equation}\label{Eqs0A00501}
\begin{split}
\boldsymbol{G}^H\boldsymbol{\bar{F}}^H\boldsymbol{\bar{F}}\boldsymbol{G} = {M_0}\boldsymbol{\bar {G}}^H\boldsymbol{{I}}_{K}\boldsymbol{\bar {G}}= {M_0}\boldsymbol{\bar {G}}^H\boldsymbol{\bar {G}}.
\end{split}
\end{equation}
Based on (\ref{Eqs0A00502}) and (\ref{Eqs0A00501}), we can obtain
\begin{equation}\label{Eqs0A005}
\begin{split}
\boldsymbol{G}^H\boldsymbol{\bar{F}}^H\boldsymbol{\bar{F}}\boldsymbol{G} = {M_0}\boldsymbol{ {G}}^H\boldsymbol{ {G}} .
\end{split}
\end{equation}
By substituting (\ref{Eqs0A005}) into (\ref{Eqs0A002}), we have
\begin{equation}\label{Eqs0A006}
\begin{split}
  \tilde{p}\left(\boldsymbol{v}\right) & = \frac{P}{M_0M}{\boldsymbol{v}^H\mathbb{E}\left[\boldsymbol{G}^H\boldsymbol{\bar{F}}^H\boldsymbol{\bar{F}}\boldsymbol{G}\right]\boldsymbol{v}} + \sigma^2 \\
  & = \frac{P}{M}{\boldsymbol{v}^H\mathbb{E}\left[\boldsymbol{G}^H\boldsymbol{G}\right]\boldsymbol{v}} + \sigma^2.
\end{split}
\end{equation}
As ${\boldsymbol{R}} = \frac{P}{M}\mathbb{E}\left[{\boldsymbol{G}}^H\boldsymbol{G}\right]$, (\ref{Eqs0A006}) can be equivalently expressed as
\begin{equation}\label{Eqs0A007}
\begin{split}
  \tilde{p}\left(\boldsymbol{v}\right) = {\boldsymbol{v}^H{\boldsymbol{R}}\boldsymbol{v}} + \sigma^2.
\end{split}
\end{equation}
The proof is thus completed.

\subsection{Proof of Lemma \ref{lemma1}}\label{app002}
   First, we prove Lemma \ref{lemma1} in the case of $K' \geq K$. By applying the spectral decomposition to the Hermitian and PSD matrix $\boldsymbol{R}$, we can express the channel autocorrelation matrix $\boldsymbol{R}$ as
   \begin{equation}\label{Eqs00305}
   \boldsymbol{R} = \sum_{k=1}^{K}{\lambda_k\boldsymbol{b}_k\boldsymbol{b}^H_k},
   \end{equation}
   where $\lambda_k > 0$ denotes the $k$-th eigenvalue and $\boldsymbol{b}_k$ denotes the $k$-th eigenvector. By defining $\boldsymbol{a}_k=\sqrt{\lambda_k}\boldsymbol{b}_k$, we have
   \begin{equation}\label{Eqs00305001}
   \boldsymbol{R} = \sum_{k=1}^{K}{\boldsymbol{a}_k\boldsymbol{a}^H_k}.
   \end{equation}
   The proof is thus completed for $K'=K$.

   While for the case of $K' > K$, we can simply rewrite $\boldsymbol{a}_1\boldsymbol{a}^H_1$ in (\ref{Eqs00305001}) as
   \begin{equation}\label{Eqs0030501}
     \boldsymbol{a}_1\boldsymbol{a}^H_1 = \sum_{j=1}^{J}{\lambda_{1,j}\boldsymbol{b}_1\boldsymbol{b}^H_1},
   \end{equation}
    where ${\lambda_{ 1, j }}=\frac{\lambda_1}{J}, 1 \leq j \leq J$ and $J = K'-K+1> 1$. By substituting (\ref{Eqs0030501}) into (\ref{Eqs00305001}), we have
    \begin{equation}\label{Eqs00305002}
       \boldsymbol{R} = \sum_{j=1}^{J}{\lambda_{1,j}\boldsymbol{b}_1\boldsymbol{b}^H_1} + \sum_{k=2}^{K}{\boldsymbol{a}_k\boldsymbol{a}^H_k} = \sum_{k=1}^{K'}{\boldsymbol{a}'_k\boldsymbol{a}'^H_k},
       \end{equation}
    with
    \begin{equation*}
      \boldsymbol{a}'_k= \left\{ {\begin{array}{*{20}{c}}
    \sqrt{\lambda_{1,k}}\boldsymbol{b}_1, & \text{if} & 1 \leq k \leq J, \\
    \boldsymbol{a}_{(k-J+1)}, & \text{if} & J < k \leq K'.
    \end{array}} \right.
    \end{equation*}
    Thus, the proof is completed for $K' > K$.

   Finally, for the case of $K' < K$, suppose that $\boldsymbol{R}$ can be expressed as
   \begin{equation}\label{Eqs00306}
     \boldsymbol{R} = \sum_{k=1}^{K'}\boldsymbol{a}_k\boldsymbol{a}^H_k.
   \end{equation}
   According to the matrix rank inequality, the rank of the summation of matrices is upper-bounded by the sum of their individual ranks, i.e.,
   \begin{equation}\label{Eqs000307}
     K=\text{rank}(\boldsymbol{R}) \le \sum_{k=1}^{K'} \text{rank}(\boldsymbol{a}_k\boldsymbol{a}^H_k) = K',
   \end{equation}
    which contradicts the precondition $K'<K$. Therefore, $\boldsymbol{R}$ cannot be expressed as (\ref{Eqs00306}) if $K'<K$. The proof is thus completed.

\subsection{Proof of Lemma \ref{lemma2}}\label{app003}
By substituting (\ref{Eqs001013}) into (\ref{Eqs003012}) and leveraging $\boldsymbol{u}=\left[{\Re}\left(\boldsymbol{v}^T\right), {\Im}\left(\boldsymbol{v}^T\right)\right]^T$ and $\boldsymbol{w}_{k} = \boldsymbol{w}_{k,1} + \jmath \boldsymbol{w}_{k,2}$, we obtain
\begin{equation}\label{Eqs0001016}
  \hat p(\boldsymbol{v}) = \sum_{k=1}^{K'}{\left|\boldsymbol{v}^H\boldsymbol{w}_k\right|^2}=\boldsymbol{v}^H\left(\sum_{k=1}^{K'}{\boldsymbol{w}_k\boldsymbol{w}_k^H}\right)\boldsymbol{v}.
\end{equation}
If ${\hat p}\left(\boldsymbol{v}\right) = {\boldsymbol{v}}^H{\boldsymbol{R}}{\boldsymbol{v}}$ holds for any $\boldsymbol{v}$, then $\boldsymbol{v}^H\left(\sum_{k=1}^{K'}{\boldsymbol{w}_k\boldsymbol{w}_k^H}- \boldsymbol{R}\right)\boldsymbol{v} = 0$ should hold for any $\boldsymbol{v}$. In this case, (\ref{Eqs001015}) must hold.

\bibliographystyle{IEEEtran}

\bibliography{References}

\end{document}